\documentclass[aps,prd,floatfix,groupedaddress,nofootinbib,superscriptaddress,twocolumn]{revtex4-1}
\usepackage{textcomp}
\usepackage{bm,color}
\usepackage{verbatim}
\usepackage{amsmath,amssymb}
\usepackage{ulem}
\definecolor{linkcolor}{rgb}{0.6,0,0}
\definecolor{citecolor}{rgb}{0,0,0.75}
\definecolor{urlcolor}{rgb}{0.12,0.46,0.7}
\usepackage[breaklinks, colorlinks, urlcolor=urlcolor, linkcolor=linkcolor,citecolor=citecolor,pdfencoding=auto]{hyperref}
\usepackage{booktabs}
\usepackage{bigstrut}
\usepackage{xspace}
\usepackage{multirow}
\usepackage{amsmath}
\usepackage{graphicx}
\usepackage{threeparttable}
\usepackage{subfigure}
\usepackage{enumitem}

\newcommand{\sptpol}{{SPTpol}}

\newcommand{\nver}{\hat{\mathbf{n}}}

\newcommand*\Bell{\ensuremath{\boldsymbol\ell}}

\newcommand{\vanish}[1]{}

\newcommand{\planck}{\textit{Planck}\xspace}
\newcommand{\sqdeg}{$\deg^2$}

\newcommand{\ACBamp}{\ensuremath{-0.049 \pm 0.096}}
\newcommand{\ACBmcmc}{\ensuremath{0.10}}
\newcommand{\ACBmcmcsqrad}{\ensuremath{0.033}}
\newcommand{\Bmcmc}{\ensuremath{17}}
\newcommand{\gagammamcmc}{\ensuremath{4.0\times 10^{-2}}}
\newcommand{\chisquareauto}{\ensuremath{7.7}}
\newcommand{\pteauto}{\ensuremath{76.5}}
\newcommand{\chisquarecross}{\ensuremath{9.8}}
\newcommand{\ptecross}{\ensuremath{55.8}}

\newcommand{\beq}{\begin{equation}}
\newcommand{\eeq}{\end{equation}}
\newcommand{\bea}{\begin{eqnarray}}
\newcommand{\eea}{\end{eqnarray}}

\begin{document}

\title{Searching for Anisotropic Cosmic Birefringence with Polarization Data from \sptpol{}}

%\shortauthors{F.~Bianchini, W.~L.~K.~Wu, et al.}
\author{F.~Bianchini}\email{fbianchini@unimelb.edu.au} \affiliation{School of Physics, University of Melbourne, Parkville, VIC 3010, Australia}
\author{W.~L.~K.~Wu} \affiliation{Kavli Institute for Cosmological Physics, University of Chicago, 5640 South Ellis Avenue, Chicago, IL, USA 60637}
\author{P.~A.~R.~Ade} \affiliation{Cardiff University, Cardiff CF10 3XQ, United Kingdom}
\author{A.~J.~Anderson} \affiliation{Fermi National Accelerator Laboratory, MS209, P.O. Box 500, Batavia, IL 60510}
\author{J.~E.~Austermann} \affiliation{NIST Quantum Devices Group, 325 Broadway Mailcode 817.03, Boulder, CO, USA 80305} \affiliation{Department of Physics, University of Colorado, Boulder, CO, USA 80309}
\author{J.~S.~Avva} \affiliation{Department of Physics, University of California, Berkeley, CA, USA 94720}
\author{L.~Balkenhol} \affiliation{School of Physics, University of Melbourne, Parkville, VIC 3010, Australia}
\author{E.~Baxter} \affiliation{Institute for Astronomy, University of Hawai'i, 2680 Woodlawn Drive, Honolulu, HI 96822, USA}
\author{J.~A.~Beall} \affiliation{NIST Quantum Devices Group, 325 Broadway Mailcode 817.03, Boulder, CO, USA 80305}
\author{A.~N.~Bender} \affiliation{High Energy Physics Division, Argonne National Laboratory, 9700 S. Cass Avenue, Argonne, IL, USA 60439} \affiliation{Kavli Institute for Cosmological Physics, University of Chicago, 5640 South Ellis Avenue, Chicago, IL, USA 60637}
\author{B.~A.~Benson} \affiliation{Fermi National Accelerator Laboratory, MS209, P.O. Box 500, Batavia, IL 60510} \affiliation{Kavli Institute for Cosmological Physics, University of Chicago, 5640 South Ellis Avenue, Chicago, IL, USA 60637} \affiliation{Department of Astronomy and Astrophysics, University of Chicago, 5640 South Ellis Avenue, Chicago, IL, USA 60637}
\author{L.~E.~Bleem} \affiliation{High Energy Physics Division, Argonne National Laboratory, 9700 S. Cass Avenue, Argonne, IL, USA 60439} \affiliation{Kavli Institute for Cosmological Physics, University of Chicago, 5640 South Ellis Avenue, Chicago, IL, USA 60637}
\author{J.~E.~Carlstrom} \affiliation{Kavli Institute for Cosmological Physics, University of Chicago, 5640 South Ellis Avenue, Chicago, IL, USA 60637} \affiliation{Department of Physics, University of Chicago, 5640 South Ellis Avenue, Chicago, IL, USA 60637} \affiliation{High Energy Physics Division, Argonne National Laboratory, 9700 S. Cass Avenue, Argonne, IL, USA 60439} \affiliation{Department of Astronomy and Astrophysics, University of Chicago, 5640 South Ellis Avenue, Chicago, IL, USA 60637} \affiliation{Enrico Fermi Institute, University of Chicago, 5640 South Ellis Avenue, Chicago, IL, USA 60637}
\author{C.~L.~Chang} \affiliation{Kavli Institute for Cosmological Physics, University of Chicago, 5640 South Ellis Avenue, Chicago, IL, USA 60637} \affiliation{High Energy Physics Division, Argonne National Laboratory, 9700 S. Cass Avenue, Argonne, IL, USA 60439} \affiliation{Department of Astronomy and Astrophysics, University of Chicago, 5640 South Ellis Avenue, Chicago, IL, USA 60637}
\author{P.~Chaubal} \affiliation{School of Physics, University of Melbourne, Parkville, VIC 3010, Australia}
\author{H.~C.~Chiang} \affiliation{Department of Physics, McGill University, 3600 Rue University, Montreal, Quebec H3A 2T8, Canada} \affiliation{School of Mathematics, Statistics \& Computer Science, University of KwaZulu-Natal, Durban, South Africa}
\author{T.~L.~Chou} \affiliation{Department of Physics, University of Chicago, Chicago, IL, USA 60637}
\author{R.~Citron} \affiliation{University of Chicago, 5640 South Ellis Avenue, Chicago, IL, USA 60637}
\author{C.~Corbett~Moran} \affiliation{TAPIR, Walter Burke Institute for Theoretical Physics, California Institute of Technology, 1200 E California Blvd, Pasadena, CA, USA 91125}
\author{T.~M.~Crawford} \affiliation{Kavli Institute for Cosmological Physics, University of Chicago, 5640 South Ellis Avenue, Chicago, IL, USA 60637} \affiliation{Department of Astronomy and Astrophysics, University of Chicago, 5640 South Ellis Avenue, Chicago, IL, USA 60637}
\author{A.~T.~Crites} \affiliation{Kavli Institute for Cosmological Physics, University of Chicago, 5640 South Ellis Avenue, Chicago, IL, USA 60637} \affiliation{Department of Astronomy and Astrophysics, University of Chicago, 5640 South Ellis Avenue, Chicago, IL, USA 60637} \affiliation{California Institute of Technology, MS 249-17, 1216 E. California Blvd., Pasadena, CA, USA 91125}
\author{T.~de~Haan} \affiliation{Department of Physics, University of California, Berkeley, CA, USA 94720} \affiliation{Physics Division, Lawrence Berkeley National Laboratory, Berkeley, CA, USA 94720}
\author{M.~A.~Dobbs} \affiliation{Department of Physics, McGill University, 3600 Rue University, Montreal, Quebec H3A 2T8, Canada} \affiliation{Canadian Institute for Advanced Research, CIFAR Program in Gravity and the Extreme Universe, Toronto, ON, M5G 1Z8, Canada}
\author{W.~Everett} \affiliation{Department of Astrophysical and Planetary Sciences, University of Colorado, Boulder, CO, USA 80309}
\author{J.~Gallicchio} \affiliation{Kavli Institute for Cosmological Physics, University of Chicago, 5640 South Ellis Avenue, Chicago, IL, USA 60637} \affiliation{Harvey Mudd College, 301 Platt Blvd., Claremont, CA 91711}
\author{E.~M.~George} \affiliation{European Southern Observatory, Karl-Schwarzschild-Str. 2, 85748 Garching bei M\"{u}nchen, Germany} \affiliation{Department of Physics, University of California, Berkeley, CA, USA 94720}
\author{A.~Gilbert} \affiliation{Department of Physics, McGill University, 3600 Rue University, Montreal, Quebec H3A 2T8, Canada}
\author{N.~Gupta} \affiliation{School of Physics, University of Melbourne, Parkville, VIC 3010, Australia}
\author{N.~W.~Halverson} \affiliation{Department of Astrophysical and Planetary Sciences, University of Colorado, Boulder, CO, USA 80309} \affiliation{Department of Physics, University of Colorado, Boulder, CO, USA 80309}
\author{J.~W.~Henning} \affiliation{High Energy Physics Division, Argonne National Laboratory, 9700 S. Cass Avenue, Argonne, IL, USA 60439} \affiliation{Kavli Institute for Cosmological Physics, University of Chicago, 5640 South Ellis Avenue, Chicago, IL, USA 60637}
\author{G.~C.~Hilton} \affiliation{NIST Quantum Devices Group, 325 Broadway Mailcode 817.03, Boulder, CO, USA 80305}
\author{G.~P.~Holder} \affiliation{Astronomy Department, University of Illinois at Urbana-Champaign, 1002 W. Green Street, Urbana, IL 61801, USA} \affiliation{Department of Physics, University of Illinois Urbana-Champaign, 1110 W. Green Street, Urbana, IL 61801, USA} \affiliation{Canadian Institute for Advanced Research, CIFAR Program in Gravity and the Extreme Universe, Toronto, ON, M5G 1Z8, Canada}
\author{W.~L.~Holzapfel} \affiliation{Department of Physics, University of California, Berkeley, CA, USA 94720}
\author{J.~D.~Hrubes} \affiliation{University of Chicago, 5640 South Ellis Avenue, Chicago, IL, USA 60637}
\author{N.~Huang} \affiliation{Department of Physics, University of California, Berkeley, CA, USA 94720}
\author{J.~Hubmayr} \affiliation{NIST Quantum Devices Group, 325 Broadway Mailcode 817.03, Boulder, CO, USA 80305}
\author{K.~D.~Irwin} \affiliation{SLAC National Accelerator Laboratory, 2575 Sand Hill Road, Menlo Park, CA 94025} \affiliation{Dept. of Physics, Stanford University, 382 Via Pueblo Mall, Stanford, CA 94305}
\author{L.~Knox} \affiliation{Department of Physics, University of California, One Shields Avenue, Davis, CA, USA 95616}
\author{A.~T.~Lee} \affiliation{Department of Physics, University of California, Berkeley, CA, USA 94720} \affiliation{Physics Division, Lawrence Berkeley National Laboratory, Berkeley, CA, USA 94720}
\author{D.~Li} \affiliation{NIST Quantum Devices Group, 325 Broadway Mailcode 817.03, Boulder, CO, USA 80305} \affiliation{SLAC National Accelerator Laboratory, 2575 Sand Hill Road, Menlo Park, CA 94025}
\author{A.~Lowitz} \affiliation{Department of Astronomy and Astrophysics, University of Chicago, 5640 South Ellis Avenue, Chicago, IL, USA 60637}
\author{A.~Manzotti} \affiliation{Kavli Institute for Cosmological Physics, University of Chicago, 5640 South Ellis Avenue, Chicago, IL, USA 60637} \affiliation{Institut d'Astrophysique de Paris, 98 bis boulevard Arago, 75014 Paris, France}
\author{J.~J.~McMahon} \affiliation{Department of Physics, University of Michigan, 450 Church Street, Ann  Arbor, MI, USA 48109}
\author{S.~S.~Meyer} \affiliation{Kavli Institute for Cosmological Physics, University of Chicago, 5640 South Ellis Avenue, Chicago, IL, USA 60637} \affiliation{Department of Physics, University of Chicago, 5640 South Ellis Avenue, Chicago, IL, USA 60637} \affiliation{Department of Astronomy and Astrophysics, University of Chicago, 5640 South Ellis Avenue, Chicago, IL, USA 60637} \affiliation{Enrico Fermi Institute, University of Chicago, 5640 South Ellis Avenue, Chicago, IL, USA 60637}
\author{M.~Millea} \affiliation{Department of Physics, University of California, Berkeley, CA, USA 94720}
\author{L.~M.~Mocanu} \affiliation{Kavli Institute for Cosmological Physics, University of Chicago, 5640 South Ellis Avenue, Chicago, IL, USA 60637} \affiliation{Department of Astronomy and Astrophysics, University of Chicago, 5640 South Ellis Avenue, Chicago, IL, USA 60637}
\author{J.~Montgomery} \affiliation{Department of Physics, McGill University, 3600 Rue University, Montreal, Quebec H3A 2T8, Canada}
\author{A.~Nadolski} \affiliation{Astronomy Department, University of Illinois at Urbana-Champaign, 1002 W. Green Street, Urbana, IL 61801, USA} \affiliation{Department of Physics, University of Illinois Urbana-Champaign, 1110 W. Green Street, Urbana, IL 61801, USA}
\author{T.~Natoli} \affiliation{Department of Astronomy and Astrophysics, University of Chicago, 5640 South Ellis Avenue, Chicago, IL, USA 60637} \affiliation{Kavli Institute for Cosmological Physics, University of Chicago, 5640 South Ellis Avenue, Chicago, IL, USA 60637} \affiliation{Dunlap Institute for Astronomy \& Astrophysics, University of Toronto, 50 St George St, Toronto, ON, M5S 3H4, Canada}
\author{J.~P.~Nibarger} \affiliation{NIST Quantum Devices Group, 325 Broadway Mailcode 817.03, Boulder, CO, USA 80305}
\author{G.~Noble} \affiliation{Department of Physics, McGill University, 3600 Rue University, Montreal, Quebec H3A 2T8, Canada}
\author{V.~Novosad} \affiliation{Materials Sciences Division, Argonne National Laboratory, 9700 S. Cass Avenue, Argonne, IL, USA 60439}
\author{Y.~Omori} \affiliation{Kavli Institute for Particle Astrophysics and Cosmology, Stanford University, 452 Lomita Mall, Stanford, CA 94305}
\author{S.~Padin} \affiliation{Kavli Institute for Cosmological Physics, University of Chicago, 5640 South Ellis Avenue, Chicago, IL, USA 60637} \affiliation{Department of Astronomy and Astrophysics, University of Chicago, 5640 South Ellis Avenue, Chicago, IL, USA 60637} \affiliation{California Institute of Technology, MS 249-17, 1216 E. California Blvd., Pasadena, CA, USA 91125}
\author{S.~Patil} \affiliation{School of Physics, University of Melbourne, Parkville, VIC 3010, Australia}
\author{C.~Pryke} \affiliation{School of Physics and Astronomy, University of Minnesota, 116 Church Street S.E. Minneapolis, MN, USA 55455}
\author{C.~L.~Reichardt} \affiliation{School of Physics, University of Melbourne, Parkville, VIC 3010, Australia}
\author{J.~E.~Ruhl} \affiliation{Physics Department, Center for Education and Research in Cosmology and Astrophysics, Case Western Reserve University, Cleveland, OH, USA 44106}
\author{B.~R.~Saliwanchik} \affiliation{Physics Department, Center for Education and Research in Cosmology and Astrophysics, Case Western Reserve University, Cleveland, OH, USA 44106} \affiliation{Department of Physics, Yale University, P.O. Box 208120, New Haven, CT 06520-8120}
\author{K.~K.~Schaffer} \affiliation{Kavli Institute for Cosmological Physics, University of Chicago, 5640 South Ellis Avenue, Chicago, IL, USA 60637} \affiliation{Enrico Fermi Institute, University of Chicago, 5640 South Ellis Avenue, Chicago, IL, USA 60637} \affiliation{Liberal Arts Department, School of the Art Institute of Chicago, 112 S Michigan Ave, Chicago, IL, USA 60603}
\author{C.~Sievers} \affiliation{University of Chicago, 5640 South Ellis Avenue, Chicago, IL, USA 60637}
\author{G.~Simard} \affiliation{Department of Physics, McGill University, 3600 Rue University, Montreal, Quebec H3A 2T8, Canada}
\author{G.~Smecher} \affiliation{Department of Physics, McGill University, 3600 Rue University, Montreal, Quebec H3A 2T8, Canada} \affiliation{Three-Speed Logic, Inc., Vancouver, B.C., V6A 2J8, Canada}
\author{A.~A.~Stark} \affiliation{Harvard-Smithsonian Center for Astrophysics, 60 Garden Street, Cambridge, MA, USA 02138}
\author{K.~T.~Story} \affiliation{Kavli Institute for Particle Astrophysics and Cosmology, Stanford University, 452 Lomita Mall, Stanford, CA 94305} \affiliation{Dept. of Physics, Stanford University, 382 Via Pueblo Mall, Stanford, CA 94305}
\author{C.~Tucker} \affiliation{Cardiff University, Cardiff CF10 3XQ, United Kingdom}
\author{K.~Vanderlinde} \affiliation{Dunlap Institute for Astronomy \& Astrophysics, University of Toronto, 50 St George St, Toronto, ON, M5S 3H4, Canada} \affiliation{Department of Astronomy \& Astrophysics, University of Toronto, 50 St George St, Toronto, ON, M5S 3H4, Canada}
\author{T.~Veach} \affiliation{Department of Astronomy, University of Maryland College Park, MD, USA 20742}
\author{J.~D.~Vieira} \affiliation{Astronomy Department, University of Illinois at Urbana-Champaign, 1002 W. Green Street, Urbana, IL 61801, USA} \affiliation{Department of Physics, University of Illinois Urbana-Champaign, 1110 W. Green Street, Urbana, IL 61801, USA}
\author{G.~Wang} \affiliation{High Energy Physics Division, Argonne National Laboratory, 9700 S. Cass Avenue, Argonne, IL, USA 60439}
\author{N.~Whitehorn} \affiliation{Department of Physics and Astronomy, University of California, Los Angeles, CA, USA 90095}
\author{V.~Yefremenko} \affiliation{High Energy Physics Division, Argonne National Laboratory, 9700 S. Cass Avenue, Argonne, IL, USA 60439}

%\correspondingauthor{Federico Bianchini}
\email{fbianchini@unimelb.edu.au}

\begin{abstract}
We present a search for anisotropic cosmic birefringence in 500 \sqdeg{} of southern sky observed at 150 GHz with the \sptpol{} camera on the South Pole Telescope. We reconstruct a map of cosmic polarization rotation anisotropies using higher-order correlations between the observed cosmic microwave background (CMB) $E$ and $B$ fields. We then measure the angular power spectrum of this map, which is found to be consistent with zero. The nondetection is translated into an upper limit on the amplitude of the scale-invariant cosmic rotation power spectrum, $L(L+1)C_L^{\alpha\alpha}/2\pi <  \ACBmcmc \times 10^{-4}$  rad$^2$ (\ACBmcmcsqrad\, \sqdeg{}, \,95\% C.L.).  This upper limit can be used to place constraints on the strength of primordial magnetic fields, $B_{1 \rm Mpc} < \Bmcmc \,{\rm nG} $ (95\% C.L.), and on the coupling constant of the Chern-Simons electromagnetic term $g_{a\gamma} < \gagammamcmc/H_I $ (95\% C.L.), where $H_I$ is the inflationary Hubble scale. For the first time, we also cross-correlate the CMB temperature fluctuations with the reconstructed rotation angle map, a signal expected to be nonvanishing in certain theoretical scenarios, and find no detectable signal. We perform a suite of systematics and consistency checks and find no evidence for contamination. 
\end{abstract}


\maketitle

%%%%%%%%%%%%%%%%%%%%%%%%
%% Introduction
%%%%%%%%%%%%%%%%%%%%%%%%
\section{Introduction} \label{sec:intro}
The exquisite mapping of the cosmic microwave background polarization (CMB) anisotropies, in particular of the odd-parity $B$-modes, is arguably the main driver of the current and upcoming experimental effort in CMB research (\citep[][SPT-3G]{bender18}; \citep[AdvACT]{henderson16}; \citep[BICEP3/Keck Array]{grayson16}; \citep[Simons Array]{suzuki16}; \citep[CLASS]{essinger14}; \citep[Simons Observatory]{simonsobservatorycollab19}; \citep[CMB-S4]{cmbs4collab19}).
Beyond providing key insights on the physics of the early universe and the large-scale matter distribution, at large ($\ell \lesssim 100$) and small ($\ell \gtrsim 100$) angular scales respectively, accurate measurements of the CMB $B$-modes open new avenues to test fundamental physics and a variety of exotic physics \citep[e.g.,][]{staggs18}.

Among the several physical processes affecting CMB photons during their cosmic journey, in this paper we focus on the cosmic birefringence (CB), i.e., the {\it in vacuo} rotation of the plane of polarization of photons over cosmological distances. 
CB naturally arises in different theoretical contexts, which can be roughly broken down into two main classes: parity-violating extensions of the standard model \citep[e.g.,][]{carroll90,pospelov09} and primordial magnetic fields (PMF, e.g., \citep{kosowsky96}).

Depending on the specific details of the physical process sourcing the cosmic polarization rotation, for example whether the underlying pseudoscalar field is homogenous or not,  we can expect a uniform rotation angle $\alpha$, an anisotropic rotation $\alpha(\nver)$ across the sky, or both.

Measurements of the constant polarization rotation angle $\alpha$ have been performed in recent years using both astrophysical sources, such as radio galaxies, and the CMB. 
So far, there has been no evidence of a nonzero uniform rotation angle $\alpha$, with statistical errors of order of 0.2$^{\circ}$ and systematic uncertainties dominating the error budget at the level of 0.3$^{\circ}$ \citep[e.g.,][]{planck15-49}.
In the absence of other foregrounds, the isotropic birefringence angle $\alpha$
is completely degenerate with a systematic error in the global orientation of
the polarization-sensitive detectors, which effectively poses an intrinsic
limiting factor in the detection of a uniform CB. 
Efforts are currently devoted
to devise strategies to improve the calibration for the polarization angle of
CMB experiments, for example using artificial calibration sources flown on drones
or balloons, using the Crab Nebula, or using the foregrounds themselves as a
calibrator see e.g., \citep{nati17,aumont19,minami19,minami20}, respectively. 

A search for an anisotropic CB effect is complementary as it is not sensitive to a systematic uniform rotation, and well-motivated, as many theoretical models predict fluctuations of the rotation angle over the sky (and many models feature a vanishing constant rotation).
The best upper limits on the amplitude of the scale-invariant anisotropic rotation power spectrum mostly come from measurements of the 4-point correlation functions in the CMB and are currently of the order $\langle (\Delta\alpha)^2\rangle^{1/2} \lesssim 0.5^{\circ}$ \citep{gluscevic12,polarbear15,bicep2keck17,contreras17,namikawa20}.
Future CMB experiments are projected to improve this limit by orders of magnitude \citep[e.g.,][]{pogosian19}.

In this paper we search for an anisotropic CB in the CMB polarization data taken with the \sptpol{} camera. 
We reconstruct a map of the rotation angle fluctuations over 500 \sqdeg{} of the southern sky and measure its angular power spectrum. 
We use this measurement to provide constraints on the amplitude $A_{\rm CB}$ of the scale-invariant cosmic rotation power spectrum $C_L^{\alpha\alpha}$ (see Sec.~\ref{sec:theory} for the definition).
Going beyond previous analyses, we also measure the cross-correlation between the reconstructed rotation angle map with the CMB temperature fluctuations $C_L^{\alpha T}$. This cross-correlation signal is expected to be nonzero in certain theoretical contexts, including some early dark energy models from the string axiverse that have recently been investigated as a possible solution to the Hubble tension \citep[e.g.,][]{caldwell11,poulin19,capparelli19}.

The paper is structured as follows.
In Sec.~\ref{sec:theory} we provide a brief overview of the main physical mechanisms that are expected to source the cosmic polarization rotation. 
We then describe the \sptpol{} dataset and simulations used in this analysis in Sec.~\ref{sec:data_and_sims}, while the details of the cosmic rotation extraction pipeline are provided in Sec.~\ref{sec:analysis}. 
We validate our analysis against systematic effects in Sec.~\ref{sec:checks}, while we present our cosmic rotation measurement and discuss its cosmological implications in Sec.~\ref{sec:results}.
Finally, we draw our conclusions in Sec.~\ref{sec:conclusions}.

%%%%%%%%%%%%%%%%%%%%%%%%
%% Theoretical Background
%%%%%%%%%%%%%%%%%%%%%%%%
\section{Theoretical Background} \label{sec:theory}
CMB polarization experiments are designed to measure the $Q$ and $U$ Stokes parameters at different locations of the sky, $\nver$. 
The presence of an anisotropic cosmic birefringence field, $\alpha(\nver)$, introduces a phase factor in the observed polarization field $\left[Q \pm iU \right](\nver) $, rotating the primordial $\tilde{Q}$ and $\tilde{U}$ Stokes parameters according to

\beq
\label{eq:pol_rot}
\left[Q \pm iU \right](\nver) = e^{\pm 2 i \alpha(\nver)}\left[\tilde{Q} \pm i\tilde{U}\right](\nver).
\eeq
Eq.~\ref{eq:pol_rot} tells us that the rotation of the CMB polarization plane breaks parity and induces an $E$-to-$B$ mixing\footnote{Similarly, a $B$-to-$E$ mixing also arises but is much smaller because the magnitude of primordial $C_{\ell}^{BB}$ is subdominant compared to $C_{\ell}^{EE}$.}  as well as a $T$-$B$ correlation since acoustic oscillations result in a nonzero $C_{\ell}^{TE}$.
As mentioned in the Introduction, we can broadly split the main physical mechanisms that could source the cosmic birefringence in two classes: parity-violating extensions of the standard model and primordial magnetic fields (PMF).

A general aspect of parity-violating scenarios is the presence of a (nearly) massless axionlike pseudoscalar field,\footnote{We can think of the axionlike field as a pseudo-Nambu-Goldstone boson (PNGB) of a spontaneously broken global $U(1)$ symmetry.} $a$, that couples to the standard electromagnetic term, $F_{\mu\nu}\tilde{F}^{\mu\nu}$, through a Chern-Simons interaction

\beq
\label{eq:lagrangian}
\mathcal{L} \supset \frac{g_{a\gamma}}{4} a F_{\mu\nu}\tilde{F}^{\mu\nu},
\eeq
where $g_{a\gamma}$ is the coupling constant which has mass-dimension $-1$, and $\tilde{F}^{\mu\nu}$ is the dual of the electromagnetic tensor.
axionlike particles naturally arise in string theory \citep[e.g.,][]{arvanitaki10,kamionkowski14} and have been discussed in the context of inflation \citep[e.g.,][]{freese90}, quintessence \citep[e.g.,][]{carroll98}, neutrino number asymmetry \citep[e.g.,][]{geng07}, baryogenesis \citep[e.g.,][]{alexander16,jimenez17}, early dark energy \citep[e.g.,][]{poulin19,capparelli19}, and dark matter \citep[e.g.,][]{gardner08,fedderke19}. See \citet{marsh16} for a review on axionlike fields in cosmology.

The Chern-Simons term in Eq.~\ref{eq:lagrangian} affects the propagation of right- and left-handed photons asymmetrically, giving rise to the phenomenon of birefringence. 
The amount of rotation is dictated by the change of the field integrated over the photon trajectory  $\Delta a$  and is given by

\beq
\label{eq:delta_a}
\alpha = \frac{g_{a\gamma}}{2}\Delta a.
\eeq
If the pseudoscalar field fluctuates over space and time, $\delta a(\nver,t)$, then anisotropies in the rotation angle $\alpha$ will also  be generated. 
For example, if $a$ is effectively a massless scalar field during inflation, the large-scale limit of the expected cosmic rotation power spectrum is \citep{caldwell11}

\beq
\label{eq:parity_claa}
\sqrt{\frac{L(L+1)C_L^{\alpha\alpha}}{2\pi}} = \frac{H_I g_{a\gamma}}{4\pi},
\eeq
where $H_I$ is the value of the Hubble parameter during the inflationary era.
The inflationary Hubble scale is related to the tensor-to-scalar ratio $r$ through $H_I = 2\pi M_{\rm pl}\sqrt{A_s r/8}\simeq \sqrt{4r} \times 10^{14} $ GeV, where  $M_{\rm pl} \simeq 2 \times 10^{18}$ GeV is the reduced Planck mass and $A_s \simeq 2.2 \times  10^{-9}$ is the primordial scalar perturbation amplitude \citep{marsh16}.

The second main mechanism that might generate cosmic birefringence is the Faraday rotation that CMB photons can undergo when passing through ionized regions permeated by a magnetic field \citep{kosowsky96}.
A PMF present at and just after last scattering would induce a rotation angle along the line-of-sight $\nver$ given by \citep[e.g.,][]{harari97}:

\beq
%\label{eq:}
\alpha(\nver) = \frac{3}{16\pi^2 e \nu^2} \int \, d{\bf l} \cdot \dot{\tau} {\bf B},
\eeq
where $ \dot{\tau}$ is the differential optical depth, ${\bf B}$ is the comoving magnetic field strength and $\nu$ is the observed frequency.

Magnetic fields are ubiquitous in the universe: they are observed in stars, low- and high-$z$ galaxies, galaxy clusters, as well as in filaments, and have typical strengths of the order of few-to-tens of $\mu$G (see \citep{widrow11,ryu12} for reviews).
While dynamo and compression amplification mechanisms are currently hypothesized to be responsible for the observed magnetic fields, they still require the presence of an initial nonzero magnetic ``seed" field. 
The specific details of the generation of such PMFs are still unclear but the main candidates mechanisms include inflationary scenarios, phase transitions, or other physical processes  (see \citep{durrer13} and references therein). 
An improved constraint on the strength of a PMF would therefore help discriminating among different early-universe scenarios.

The simplest proposed inflationary models of magnetogenesis predict a scale-invariant PMF \citep[e.g.,][]{turner88,ratra92}, which results in a scale-invariant cosmic rotation power spectrum \citep{de13,pogosian14}:

\beq
\label{eq:pmf}
\sqrt{\frac{L(L+1)C_L^{\alpha\alpha}}{2\pi}} = 1.9 \times 10^{-4} \left( \frac{\nu}{150 {\rm GHz}}\right)^{-2} \left( \frac{B_{1\rm Mpc}}{1~{\rm nG}}\right).
\eeq
Thanks to its characteristic frequency dependence, Faraday rotation can in principle be disentangled from other sources of birefringence by performing a multifrequency analysis.
Note that, in addition to the frequency-dependent $B$-modes induced by Faraday rotation, the metric perturbations and Lorentz force associated with the PMF also generate vector and tensor $B$-modes with angular spectra whose shape resembles those produced by primordial gravitational waves and lensing \citep[e.g.,][]{seshadri01,shaw10_pmf}.
Considering that these unaccounted contributions from PMF to $B$-modes can bias future constraints on inflationary gravitational waves \citep[e.g.,][]{renzi18}, a 4-point function analysis such as the one presented in this paper provides an informative cross-check on the sources of polarized $B$-modes.

Since the majority of the physical mechanisms discussed above generically predict a scale-invariant power spectrum at large scales ($L \lesssim 100$), and to facilitate a comparison with previous studies, we consider our reference power spectrum to take the following form

\beq
\label{eq:theory_cl}
\frac{L(L+1)}{2\pi} C_L^{\alpha\alpha} = A_{\rm CB} \times 10^{-4} \quad [\text{rad}^2].
\eeq
This will be used to generate Gaussian realizations of the cosmic birefringence field $\alpha(\nver)$, as discussed in Sec.~\ref{sec:sims}, and to fit the reconstructed power spectrum in Sec.~\ref{sec:results}.
From Eq.~\ref{eq:theory_cl} it is clear that the ability to map out the largest scales on the sky translates into more stringent constraints on the amplitude of the scale-invariant cosmic rotation power spectrum.

Note that here we only consider the scale-invariant cosmic rotation power spectrum that, despite being the simplest and most widely predicted one, does not cover all the possible scenarios. For example, causal PMFs tend to have very blue CB power spectra and so do axionlike models where the symmetry breaking scale is below that of inflation.

%%%%%%%%%%%%%%%%%%%%%%%%
%% Data and Sims
%%%%%%%%%%%%%%%%%%%%%%%%
\section{Data and Simulations} \label{sec:data_and_sims}
In this section we discuss the \sptpol{} dataset, the data processing, and the suite of simulated skies used in the analysis.

\subsection{\sptpol{} 500 \sqdeg{} data} \label{sec:data}
This work makes use of data at 150 GHz from the \sptpol{} camera on the South Pole Telescope. Details on the telescope and camera can be found in \cite{padin08,carlstrom11,henning12,sayre12}.

The \sptpol{} survey field is a 500 \sqdeg{} patch of the southern sky extending from 22$^h$ to 2$^h$ in right ascension (R.A.) and from -65$^{\circ}$ to -50$^{\circ}$ in declination.
In this analysis we use the same dataset employed in the CMB lensing analysis of \citet{wu19} and we refer the reader to that work for a detailed description of the data reduction. 
Here we briefly summarize the main properties of the dataset and the resulting maps. 

The dataset comprises 3491 independent observations of the 500 \sqdeg{} field taken between April 30, 2013 and October 27, 2015. 
Each observation consists of time-ordered data (TOD) for each \sptpol{} bolometer. 
TOD are filtered and calibrated relative to each other before being binned into maps. 
For every constant-elevation scan,\footnote{We define a scan as a sweep of the telescope from one side of the field to the other.} and for every bolometer, a third- or fifth-order Legendre polynomial (depending on that specific scan observing strategy) is subtracted from the TOD. 
This effectively acts as a high-pass filter to suppress atmospheric fluctuations \citep[e.g.][]{lay00}. 
TOD are additionally low-pass filtered at a frequency corresponding to an effective multipole of $\ell=7500$ to prevent aliasing at the pixelization scale.
Electrical cross-talk between detectors is also corrected at the TOD level as described in \citet{henning18}.

We calibrate the individual bolometer TOD relative to one another by using a combination of regular observations of the Galactic HII region RCW38 and an internal chopped thermal source \citep{crites15}.
The TOD are finally accumulated into $\{T,Q,U\}$ maps using the oblique Lambert azimuthal equal-area projection with square 1' $\times$ 1' pixels.

A number of corrections are applied to the coadded maps. 
We deproject the monopole $T \to P$ leakage term from the polarization $Q$ and $U$ maps by subtracting a copy of the temperature map rescaled by the following leakage factors, $\epsilon^Q=0.018$ and $\epsilon^U=0.008$. 
We also apply a global polarization rotation angle of $0.63^{\circ} \pm 0.04^{\circ}$, calibrated by minimizing the observed $TB$ and $EB$ power spectra \citep{keating12}, to rotate the $Q$ and $U$ maps. Note that by applying this self-calibration technique we lose any sensitivity to a uniform rotation angle $\alpha$, however this does not represent an issue for the current analysis since we are interested in the anisotropic component. 
The final absolute calibration $T_{\rm cal}=0.9088$ and polarization efficiency (or polarization calibration factor) $P_{\rm cal}=1.06$ are obtained by comparing \sptpol\ maps to the CMB maps produced by \planck.
The polarization efficiency $P_{\rm cal}$ is further multiplied by a multiplicative factor, 1.01 as determined in \citet{henning18}, to account for potential biases in the \planck's polarization efficiency estimate (see \citep{wu19} for details).
The calibrated temperature map is obtained by multiplying the observed map by $T_{\rm cal}$ while the calibrated polarization maps are obtained by multiplying the $Q$ and $U$ maps by $T_{\rm cal} \times P_{\rm cal}$. 

Three main effects suppress power observed in the maps: the data filtering, the telescope angular response function (or beam), and the pixelization. 
The two-dimensional (2D) \sptpol{} transfer function $F_{\Bell}^{\rm filt}$ is estimated using noise-free maps that have been processed by the mock-observing pipeline while the beam $F_{\ell}^{\rm beam}$ is measured using Venus observations as described in \citet{henning18}.\footnote{Here and throughout the paper we adopt the flat-sky approximation and indicate the wavevector in the 2D Fourier plane with $\Bell$ while $\ell$ denotes its magnitude (and is equivalent to the multipole number).}
The pixel window function  $F_{\ell}^{\rm pix}$ is the 2D Fourier transform of a square 1' pixel.
The total transfer function is thus modelled as $F_{\Bell}^{\rm tot} = F_{\Bell}^{\rm filt}F_{\ell}^{\rm beam}F_{\ell}^{\rm pix}$.

We create a boundary mask that down-weights the noisy edges of the $\{T,Q,U\}$ maps.
Additionally, we mask bright point sources with flux density greater than 6 mJy at either 95 or 150 GHz in the 500 \sqdeg{} field using a 5' radius.

The final product of the data processing consists in a set of three coadded and masked maps, $T(\nver)$, $Q(\nver)$, $U(\nver)$, at a frequency of 150 GHz. 
The noise levels calculated in the $1000 < \ell < 3000$ range are 11.9 $\mu$K-arcmin and 8.5 $\mu$K-arcmin for the coadded temperature and polarization maps respectively.\footnote{Atmospheric noise causes a higher noise level in $T$ than in $Q$ or $U$.}

\subsection{Simulations} \label{sec:sims}
This analysis relies heavily on accurate simulations of the microwave sky to calibrate noise biases, to calculate uncertainties, and to place constraints on the amplitude of the scale-invariant cosmic rotation power spectrum (see Sec.~\ref{sec:constraints}). 
We follow the approach of \citet{story15} and \citet{wu19} to create simulations that include primary CMB, foregrounds, and instrumental noise. 

We start by generating correlated realizations of the spherical harmonic coefficients $a_{\ell m}$ of the unlensed $T$, $E$, and $B$ fields, as well as the CMB lensing potential $\phi$ and anistropic rotation angle field $\alpha$, using \textsc{Healpix} \citep{gorski05}. 
The input cosmology is the best-fit $\Lambda$CDM model to the 2015 \planck \texttt{plikHM\_TT\_lowTEB\_lensing} dataset in \citet{planck15-13}. 
The CMB $a_{\ell m}$ are then transformed to maps and lensed according to the $\phi$ realizations using \textsc{LensPIX} \citep{lewis05}. 
After lensing is applied to the CMB maps, the polarization $Q$ and $U$ Stokes parameters are further rotated in real space according to Eq.~\ref{eq:pol_rot}.
The lensed and rotated $\{T,Q,U\}$ maps are then transformed back to the harmonic space where the foregrounds are added (see below) and the $a_{\ell m}$ are multiplied by the instrument beam function $F^{\rm beam}_{\ell}$.
Finally, the beam-convolved $a_{\ell m}$ coefficients are evaluated on an equidistant cylindrical projection (ECP) grid before ``mock-observing" the realizations using the pointing information from actual observations. The simulated TOD are then filtered and processed identically to actual telescope data.

The foreground components are modelled as Gaussian realizations of the underlying power spectra. 
Note that neglecting the non-Gaussian contribution, especially from polarized Galactic foregrounds, might introduce a bias in the reconstructed cosmic rotation power spectrum. 
To assess contaminations induced by non-Gaussian foregrounds we adopt a multifaceted strategy. 
As discussed in Sec.~\ref{sec:const_checks} and \ref{sec:dust}, we first investigate potential foreground contamination by varying the minimum and maximum CMB $E/B$-mode multipoles used in the reconstruction. These two tests probe the main expected sources of non-Gaussian foreground emission, namely Galactic dust at low multipoles and polarized point sources at high multipoles. We further test for contamination by Galactic dust using dedicated non-Gaussian full-sky dust $Q/U$ simulations based on the work by \cite{vansyngel18}. 
As we will demonstrate, the impact of non-Gaussian foregrounds on the measured cosmic rotation power spectrum is negligible.
Even though the main scope of this work is the analysis of polarization data, we incorporate foreground emissions relevant for both temperature and polarization.
The simulated foregrounds include the thermal and kinematic Sunyaev-Zel'dovich (tSZ and kSZ) effects, and emission from the cosmic infrared background (CIB), radio sources, and Galactic dust.
The kSZ and tSZ spectral shapes are taken from the \citet{shaw10} model, with amplitudes chosen to match the \citet{george15} results, $\mathcal{D}_{3000}^{\rm kSZ+tSZ}=5.66\,\mu$K$^2$. 
Similarly, the modelling of the clustered and shot-noise CIB components  is taken from \citet{george15}, with $\mathcal{D}_{\ell}^{\rm CIB, cl} \propto \ell^{0.8}$ and corresponding amplitudes of $\mathcal{D}_{3000}^{\rm CIB, cl} =3.46 \,\mu$K$^2$ and $\mathcal{D}_{3000}^{\rm CIB, P} =9.16 \,\mu$K$^2$.
The radio source emission is described by $\mathcal{D}_{\ell}^{\rm radio} \propto \ell^2$ and $\mathcal{D}_{3000}^{\rm radio} =1.06 \,\mu$K$^2$.  
A 2\% polarization fraction is assumed for the Poisson-distributed components of the extragalactic polarized emission \citep{gupta19}. 
The temperature and polarization Galactic dust power is modelled as power laws with $\mathcal{D}^{\rm dust}_{\ell} \propto \ell^{-0.42}$ and amplitudes given by $\mathcal{D}^{TT,\rm dust}_{80}=1.15\, \mu$K$^2$, $\mathcal{D}^{EE,\rm dust}_{80}=0.0236 \,\mu$K$^2$, and $\mathcal{D}^{BB,\rm dust}_{80}=0.0118 \,\mu$K$^2$ \citep{keisler15}.

Instrumental noise is then added to the simulated mock-observed skies through a jackknifing approach. We first take all of the observations, split them in two sets, and then subtract the coadd of one half from the coadd of the remaining half. This process is repeated for as many times as the number of simulations by randomly grouping the observations into two halves.

We generate four sets of simulations:
\begin{enumerate}[label=(\Alph*)]
\item 400 lensed simulations;
\item 400 lensed and rotated simulations (same lensed primary CMB as set A);
\item 100 lensed and rotated simulations with different realizations of the CMB but the same realizations of $\alpha$ as the first 100 simulations in Set B;
\item 100 lensed simulations (lensed primary CMB different from set B).
\end{enumerate}
Each of the  two sets of 400 skies has the same underlying lensed primary CMB foregrounds, and instrumental noise. 
The suite A, which we refer to as the ``unrotated" simulation set, does not include the effect of cosmic birefringence, while the skies in the suite B, referred to as the ``rotated" set, are rotated using Eq.~\ref{eq:pol_rot}.
The rotated simulations are used to validate our cosmic rotation quadratic estimator, while the unrotated simulations, considered to be our baseline simulation set, are used to debias the measured power spectrum and estimate its uncertainties.
The main source of bias, the disconnected $N^{(0)}_L$ bias, is measured using the entire unrotated simulation suite.
From both the A and B simulation sets, we use 100 skies to estimate the mean-field term $\bar{\alpha}^{\rm MF}$, specifically 50 simulations for each of the two rotation anisotropy estimates $\hat{\alpha}$ that enter the CB spectrum calculation (see Eq.~\ref{eq:raw_spectrum}).
The remaining 300 simulations are used to calculate the statistical uncertainties on the measured cosmic rotation power spectrum.
An additional set of 100  unrotated skies (set D) is used to estimate the lensing bias term (see Sec.~\ref{sec:spectra}). 
The $N_L^{(1)}$ bias is estimated using a different set of 100 noiseless rotated skies (set C).
These are 100 simulations of primary CMB and are lensed by 100 corresponding different Gaussian realizations of the CMB lensing field.
We subsequently split them into two groups and rotate each sky from each group using the same cosmic birefringence field $\alpha(\nver)$ (see Sec.~\ref{sec:spectra}).

%%%%%%%%%%%%%%%%%%%%%%%%
%% Methods
%%%%%%%%%%%%%%%%%%%%%%%%
\section{Analysis Framework} \label{sec:analysis}
In this section we sketch the steps to reconstruct the rotation angle anisotropies from the observed CMB polarization maps and to obtain an unbiased estimate of their power spectrum.

\subsection{Anisotropic cosmic birefringence quadratic estimator}
\label{sec:qe}
Similarly to CMB lensing, the cosmic polarization rotation breaks the statistical isotropy of the CMB polarization field, correlating previously independent multipoles across different angular scales on the sky. The induced off-diagonal mode-mode covariance can then be exploited to reconstruct the rotation angle anisotropy field $\alpha(\nver)$ by properly averaging pairs of filtered CMB maps in harmonic space \citep{kamionkowski09,yadav09,gluscevic09,namikawa17}:

\beq
\label{eq:alpha_est_bias}
\bar{\alpha}^{EB}_{\bf L} = \int d^2 \Bell \, W_{\Bell,\Bell-{\bf L}}^{\alpha, EB} \bar{E}_{\Bell}\bar{B}_{\Bell-{\bf L}}^*.
\eeq
Here, $\bar{E}$ and $\bar{B}$ are the inverse variance-filtered $E$ and $B$ fields, $\Bell$ and  $\bf L$ are the CMB and cosmic rotation Fourier modes, and $W_{\Bell,\Bell-{\bf L}}^{\alpha, EB}$ is a weight function that describes the rotation-induced mode coupling,\footnote{Note that we ignore the lensing-induced term proportional to $C_{\ell}^{BB}$ since its impact has been shown to be negligible \citep{namikawa17,bicep2keck17}.}

\beq
\label{eq:w_alpha}
W_{\Bell,\Bell-{\bf L}}^{\alpha, EB} = 2 C_{\ell}^{EE} \cos{2(\phi_{\Bell} - \phi_{{\bf L} -\Bell})},
\eeq
where $\phi_{\Bell}$ is the angle of $\Bell$ measured from the Stokes $Q$ axis.
Note that, at linear order, the cosmic birefringence weight function $W_{\Bell,\Bell-{\bf L}}^{\alpha, EB}$ is nearly orthogonal to that of CMB lensing \citep{gluscevic09}.
While in principle other quadratic combinations of the CMB fields can be formed to reconstruct the cosmic rotation (see Tab. 1 from \citep{yadav09} for the full list), here we only use the $EB$ estimator since it provides the highest sensitivity. 
Therefore we drop the $EB$ superscript for the rest of the paper.

The input CMB polarization maps are filtered with an inverse-variance (C$^{-1}$) filter to down-weight noisy modes and to increase the sensitivity to the cosmic birefringence. 
Details about the map filtering can be found in \citep{story15,wu19}.
In this analysis we only use CMB modes with $|\Bell_x|>100$ and $|\Bell | < 3000$, to account for the impact of TOD filtering and mitigate foreground contamination. 
The effect of varying the minimum and maximum CMB multipoles on the reconstructed cosmic rotation is discussed in Sec.~\ref{sec:const_checks}.

The cosmic rotation anisotropies $\bar{\alpha}_{\bf L}$ measured with Eq.~\ref{eq:alpha_est_bias} are a biased estimate of the true cosmic rotation anisotropies $\alpha_{\bf L}$ and have to be normalized by a response function $\mathcal{R}_{\bf L}$.
This response function is calculated analytically and reads:

\beq
\mathcal{R}_{\bf L} = \int d^2 \Bell \, W_{\Bell,\Bell-{\bf L}} W_{\Bell,\Bell-{\bf L}} \mathcal{F}^E_{\Bell} \mathcal{F}^{B}_{\Bell-{\bf L}},
\eeq
where $\mathcal{F}^X_{\ell} = \left (C_{\ell}^{XX} + N_{\ell}^{XX}\right)^{-1}$ describes the diagonal approximation of the inverse-variance filter applied to the input $E$ and $B$ fields. 
We estimate the deviations from the ideal response function induced by nonstationary effects such as the survey boundary and anisotropic filtering by calculating the cross-spectrum between the input and birefringence anisotropies reconstructed from the $A_{\rm CB}=1$ simulations, $\mathcal{R}_{\bf L}^{\rm MC} = \langle \hat{\alpha}_{\bf L}^{\rm sim} (\alpha_{\bf L}^{\rm in})^*\rangle/\langle |\alpha_{\bf L}^{\rm in}|^2 \rangle$.
We find that this multiplicative correction is small, $\mathcal{R}_L^{\rm MC}\lesssim 5\%$, and approximately constant across the multipole range considered here.
Instead of perturbatively correcting the normalization by applying $\mathcal{R}_L^{\rm MC}$, we marginalize over a constant rescaling factor of the response function at the likelihood level, as discussed in detail in Sec.~\ref{sec:constraints}.
This approach presents some advantages.
To better see this, consider that the amplitude of the CB power spectrum is degenerate with a multiplicative correction of the estimator's normalization, which we recall is also estimated with a degree of uncertainty itself. 
While the application of a misestimated $\mathcal{R}_L^{\rm MC}$ would still yield unbiased results in the null hypothesis case (as is the case here), this could potentially lead to small biases on the recovered $A_{\rm CB}$ constraint if there is a non-negligible amount of CB in the data.
Therefore by including $\mathcal{R}_L^{\rm MC}$ in the likelihood calculation and marginalizing over it we are effectively absorbing our ignorance of the exact $\mathcal{R}_L^{\rm MC}$ into the $A_{\rm CB}$ inference, resulting in an unbiased and robust constraint.

We further subtract a small mean-field correction $\bar{\alpha}_{\bf L}^{\rm MF}$, estimated by averaging $\bar{\alpha}$ reconstructed from many input lensed masked CMB simulations, to account for anisotropic features, such as inhomogeneous noise and mask-induced mode-coupling, which can mimic the effects of birefringence.
The final estimate of the unbiased cosmic rotation map is thus

\beq
\label{eq:alpha_est}
\hat{\alpha}_{\bf L} = \mathcal{R}_{\bf L}^{-1} \left (\bar{\alpha}_{\bf L} - \bar{\alpha}_{\bf L}^{\rm MF} \right).
\eeq

\subsection{Power spectrum estimation}
\label{sec:spectra}
The raw cosmic rotation power spectrum $C_L^{\hat{\alpha}\hat{\alpha}}$ can be measured by correlating the cosmic birefringence map $\hat{\alpha}_{\bf L}$ obtained with Eq.~\ref{eq:alpha_est} with itself:

\beq
\label{eq:raw_spectrum}
C_L^{\hat{\alpha}\hat{\alpha}} \equiv f_{\rm mask}^{-1}\sum_{|{\bf L}| = L} \langle \hat{\alpha}_{\bf L}\hat{\alpha}_{\bf L}^*  \rangle,
\eeq
where $f_{\rm mask}$ is the average value of the fourth power of the fiducial mask.
The cosmic rotation estimator is quadratic in the CMB fields, therefore its power spectrum probes the four-point correlation function of the CMB anisotropies. 
Eq.~\ref{eq:raw_spectrum} is a biased estimate of the true cosmic rotation power spectrum $C_L^{\alpha\alpha}$ and must be corrected for a number of bias terms.

The most significant contribution to the noise budget comes from the disconnected, or Gaussian, $N^{(0)}_L$ bias. 
This term arises from chance correlations in the primary CMB, foregrounds, and noise, hence it is present even in the absence of CB.
To accurately estimate this contribution we use the realization-dependent algorithm introduced in \citet{namikawa13} which reduces the sensitivity to the mismatch between the observed and simulated CMB fluctuations and suppresses the covariance between bandpowers:\footnote{We have omitted the $\alpha\alpha$ superscript for clarity.}

\beq
N^{(0),\rm RD}_L = \langle 4 \hat{C}_L^{di} - 2 \hat{C}_L^{ij} \rangle.
\eeq
Here $\hat{C}_L^{di}$ denotes a spectrum where one leg\footnote{Here ``leg" denotes one of the two CMB fields entering the quadratic estimator.} of the quadratic estimator is fixed to be the data and the second leg is simulation $i$, $\hat{C}_L^{ij}$ is the cross-spectrum between two simulations with $j = i+1$ (cyclically), and the angle brackets denote the average over simulations.

Even after subtracting the disconnected bias, there exists a non-negligible correction from the lensing-induced trispectrum \citep{namikawa17}.
We estimate the lensing bias by subtracting $N^{(0)}_L$  from the power spectrum of a different set of unrotated simulations:\footnote{The standard $N^{(0)}_L$ bias used here can be estimated from simulations as $N^{(0)}_L = \langle 2\hat{C}_L^{ij} \rangle$.}

\beq
N^{\rm lens}_L = \langle \hat{C}_L^{ii} - N^{(0)}_L \rangle.
\eeq
From the rotated simulations we further subtract the connected bias, known as $N^{(1)}_L$ because it is first order in $C_L^{\alpha\alpha}$, which we estimate as follows \citep{story15}:

\beq
N^{(1)}_L = \langle 2 \hat{C}_L^{ii'} - 2 \hat{C}_L^{ij} \rangle,
\eeq
where $\hat{C}_L^{ii'}$ is the power spectrum constructed from two sets of simulations that share the same input CB field $\alpha$ but different lensed CMB (see Sec.~\ref{sec:sims}).

\begin{figure}
	\includegraphics[width=\columnwidth]{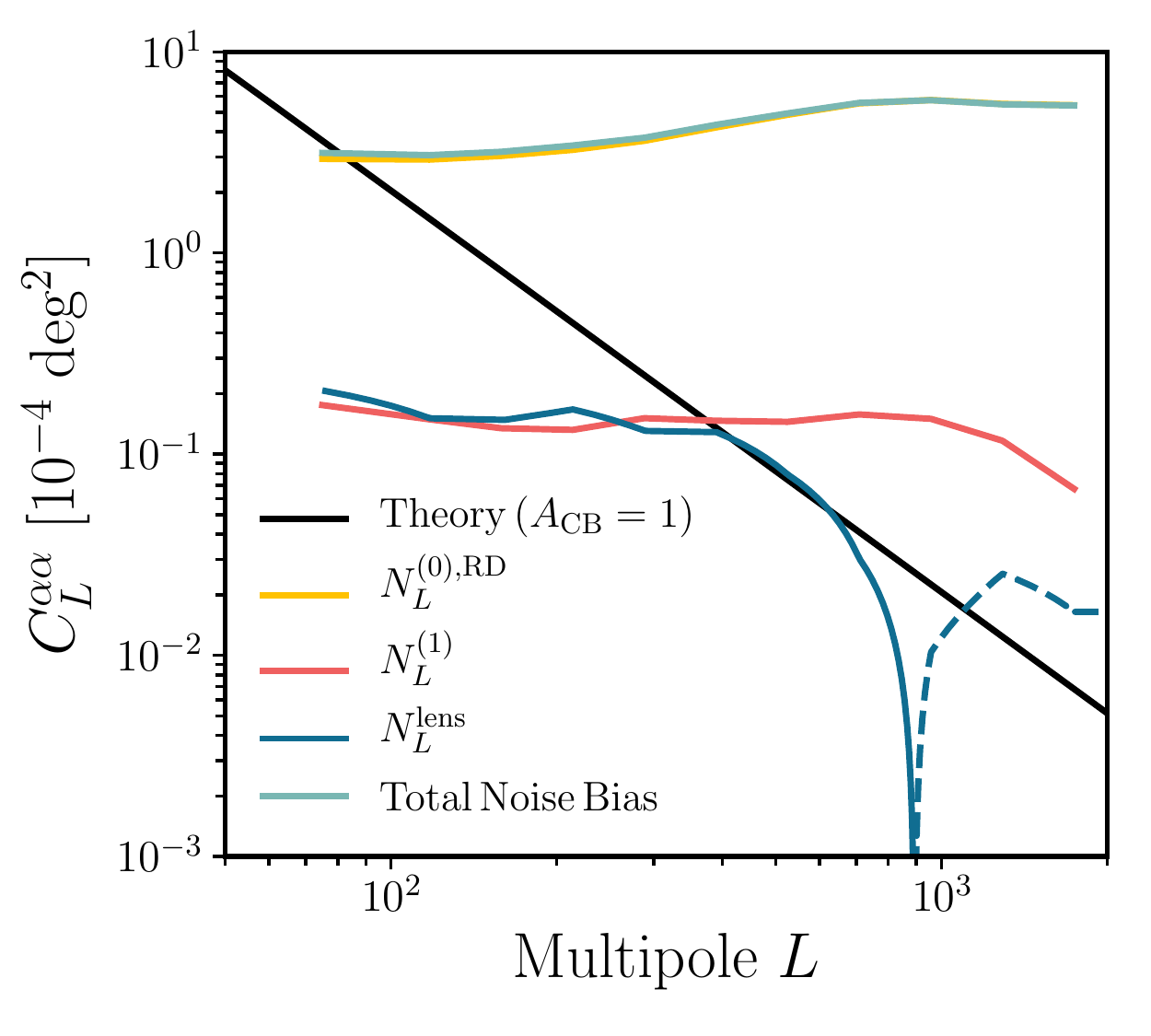}
    \caption{Noise biases for the cosmic rotation reconstruction. The theoretical scale-invariant cosmic rotation power spectrum with unit amplitude ($A_{\rm CB}=1$) is shown by the black solid line. The main source of noise, the Gaussian $N_L^{(0),{\rm RD}}$ bias, is shown by the yellow solid line and is estimated with the realization-dependent approach. The blue solid (dashed) line shows the positive (negative) values of the lensing bias $N_L^{\rm lens}$. The sum of $N_L^{(0),{\rm RD}}$ and $N_L^{\rm lens}$ is the total noise bias (cyan solid line) that we subtract from the measured raw power spectrum $C_L^{\hat{\alpha}\hat{\alpha}}$. For reference, the $N^{(1)}_L$ bias (calculated for $A_{\rm CB}=1$ and not subtracted from the observed spectrum) is shown by the red solid line. See the text for further details.}
    \label{fig:biases}
\end{figure}

The final unbiased estimate of the cosmic rotation power spectrum is thus

\beq
\hat{C}^{\alpha\alpha}_L = C_L^{\hat{\alpha}\hat{\alpha}} - N_L^{(0),\rm RD}  - N_L^{\rm lens} - N_L^{(1)}.
\eeq
We stress once again that the $N^{(1)}_L$ bias term is removed from the rotated simulations but not from the unrotated ones and, most importantly, not from the data since we are agnostic about the presence of cosmic rotation. 
Fig.~\ref{fig:biases} shows the relative magnitude of the various bias terms in our  analysis.

\subsection{Binned spectrum and amplitude}
\label{sec:bin_amp}
We measure the cosmic rotation power spectrum in 11 multipole bins in the range $50 \le L \le 2000$.
We refer to these binned power spectrum values as ``bandpowers."
We first estimate the per-bin amplitude by taking the ratio between the de-biased cosmic rotation spectrum and the input theory spectrum

\beq
A_b \equiv \frac{\hat{C}_b^{\alpha\alpha}}{C_b^{\alpha\alpha,\rm theory}},
\eeq
where $b$ stands for a binned quantity.
$C_b$ is the weighted average of $C_{\bf L}$ (either theory or data) within each bin 

\beq
C_b = \frac{\sum_{{\bf L }\in b} w_{\bf L} C_{\bf L}}{\sum_{{\bf L} \in b} w_{\bf L}},
\eeq
where the weights $w_{\bf L} = C_L^{\alpha\alpha,\rm theory}/{\rm Var}(C_L^{\hat{\alpha}\hat{\alpha}})$ are chosen to maximize the signal-to-noise and ${\rm Var}(C_L^{\hat{\alpha}\hat{\alpha}})$ is estimated from unrotated simulations. 
The overall cosmic rotation amplitude $A_{\rm CB}$ is obtained similarly to the bin-by-bin amplitude but extending the summation over the whole $L$ range.

Finally, the reported bandpowers are calculated as the product of the recovered amplitude and the input theory at the bin center $L_b$,

\beq
\hat{C}_{L_b}^{\alpha\alpha} \equiv A_b C_{L_b}^{\alpha\alpha,\rm theory}.
\eeq

\begin{figure}
	\includegraphics[width=\columnwidth]{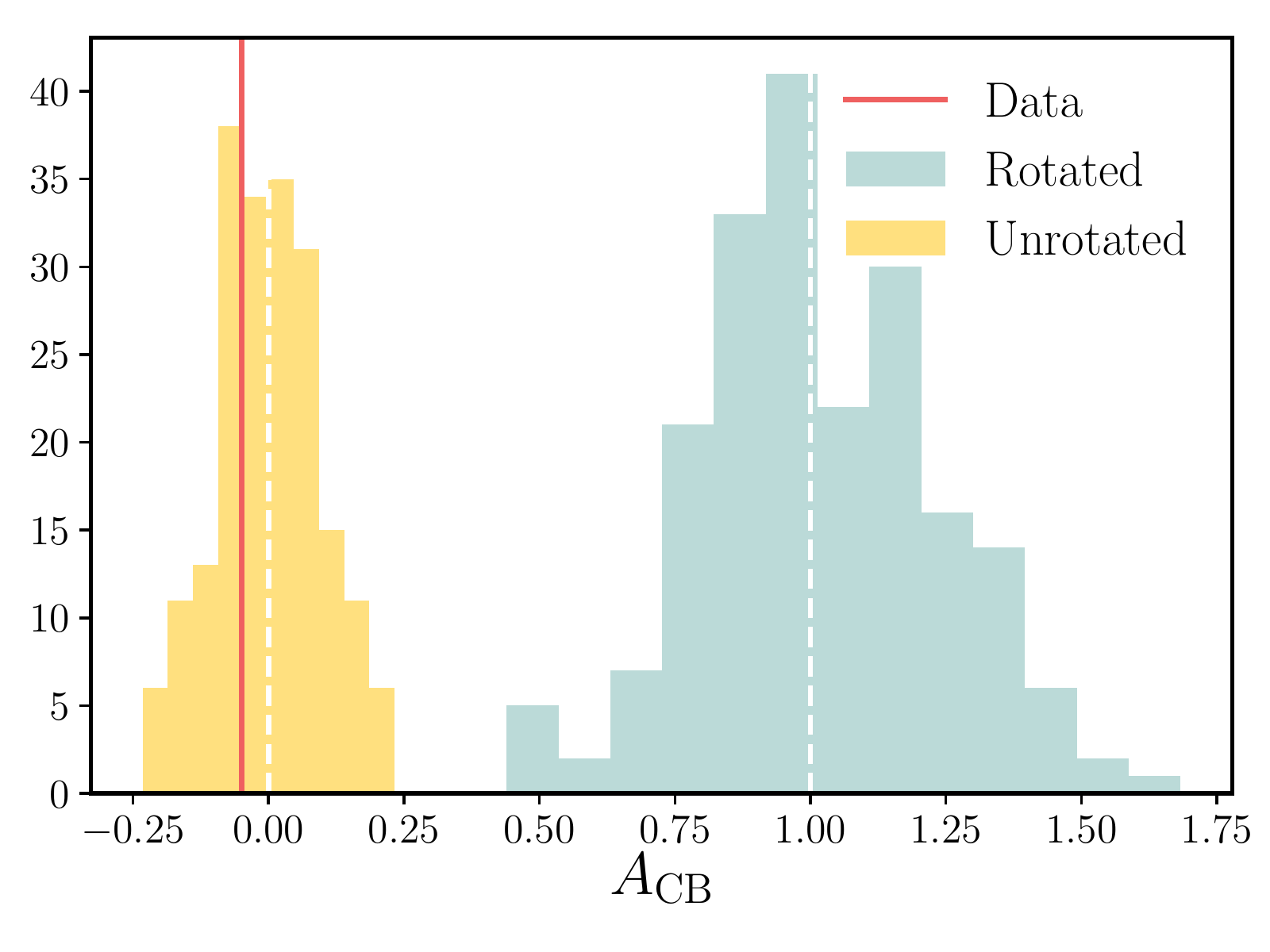}
    \caption{Distribution of the reconstructed amplitudes $A_{\rm CB}$ of the scale-invariant cosmic rotation power spectrum from unrotated (yellow histogram) and rotated (light green histogram) simulations. The corresponding $A_{\rm CB}$ value found from the observed data is shown by the red vertical line.}
    \label{fig:amps}
\end{figure}

The distribution of the recovered scale-invariant CB spectrum amplitudes from rotated and unrotated simulations is shown in Fig.~\ref{fig:amps} by the light green and yellow histograms respectively.

%

%%%%%%%%%%%%%%%%%%%%%%%
%% Systematics
%%%%%%%%%%%%%%%%%%%%%%%
\section{Analysis Validation} \label{sec:checks}
In this section we perform a suite of consistency checks and systematic tests to  validate the robustness of the results presented here.

\subsection{Consistency Checks}\label{sec:const_checks}
For each check we vary one aspect of the analysis and rerun the whole reconstruction pipeline to obtain $\hat{C}_{L_b}^{\alpha\alpha, \rm sys}$ from the data and from the set of simulations.
To assess the consistency between different analysis variations we calculate two summary statistics.
Specifically, we measure the difference between the bandpowers obtained from the baseline and modified analyses, $\Delta \hat{C}_{L_b}^{\alpha\alpha}= \hat{C}_{L_b}^{\alpha\alpha} - \hat{C}_{L_b}^{\alpha\alpha, \rm sys}$, as well as the corresponding amplitude-difference, $\Delta A_{\rm CB} = A_{\rm CB} - A_{\rm CB}^{\rm sys}$.
Both the bandpower- and amplitude-differences are then compared to the distributions inferred from the unrotated simulations.

\begin{figure}
	\includegraphics[width=\columnwidth]{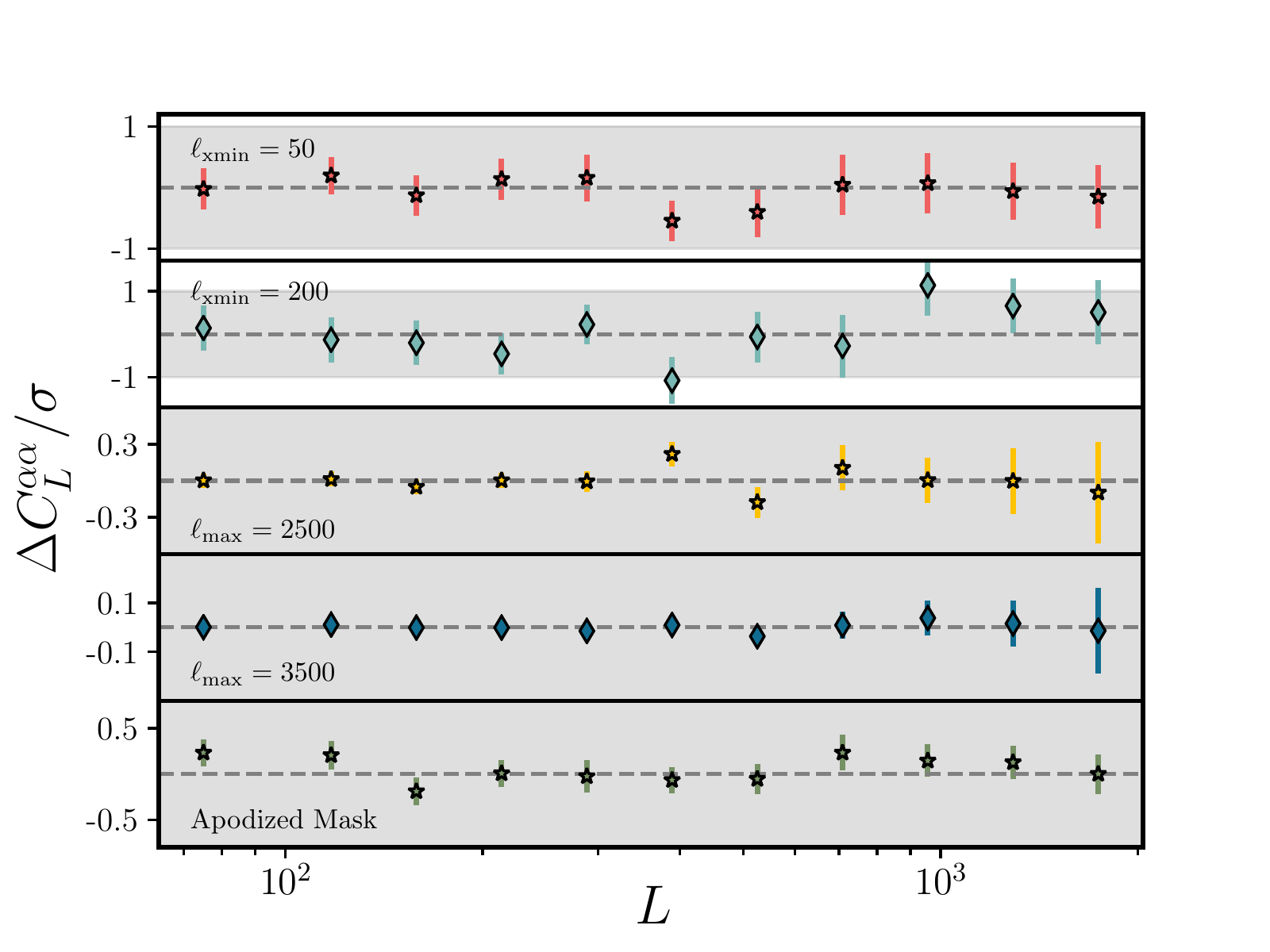}
    \caption{Consistency tests summary. The difference bandpower $\Delta\hat{C}_{L_b}^{\alpha\alpha}$ between the baseline/alternate analyses and their uncertainties are scaled by the $1\,\sigma$ cosmic rotation uncertainties in that specific bin. The grey shaded regions indicate the $1\,\sigma$ uncertainties on the baseline measurement of $\hat{C}_L^{\alpha\alpha}$. As can be seen, the induced shifts are generally only a small fraction of the statistical bandpower uncertainties.}
    \label{fig:claa_ratio_syst}
\end{figure}

The first metric quantitatively assesses the consistency by calculating the $\chi^2$ of the data difference-spectrum against the mean found in simulations using the variance of the simulation difference-spectra $\sigma_{b, \rm sys}^2$ as the uncertainty:

\begin{equation}
\chi^2_{\rm sys} = \sum_b \frac{\left( \Delta \hat{C}_{L_b}^{\alpha\alpha} - \langle \Delta \hat{C}_{L_b}^{\alpha\alpha, \rm sim}\rangle \right)^2}{\sigma_{b, \rm sys}^2}.
\end{equation}
The  probability-to-exceed (PTE) of the above $\chi^2$ is then calculated directly from simulations as the percentage of simulations that have a $\chi^2$ larger than that found for the data. 
In Fig.~\ref{fig:claa_ratio_syst} we provide a visual summary of these bandpower-difference tests.
Note that both the induced shifts and their uncertainties are only a small fraction of the statistical bandpower uncertainties. 

The second metric compares instead the shift induced by the analysis variation on the inferred cosmic rotation amplitude $\Delta A_{\rm CB}$ to the variance of the simulation difference-amplitudes $\sigma(\Delta A_{\rm CB})$. 
In a similar fashion to the bandpower-difference case, the PTE  is calculated from simulations as the percentage of simulations that have a difference-amplitude with a larger magnitude than $\Delta A_{\rm CB}$ for the data. 

The $\chi^2$ and PTEs from the different tests are listed in Tab.~\ref{tab:const_check}. 
As can be seen, the analysis variations produce bandpowers and cosmic rotation amplitudes consistent with the ones found in the baseline analysis.\\

\textit{Varying $\ell_{\rm xmin}$,  $\ell_{\rm max}$}: By varying the multipole range of the input $E$- and $B$-mode maps we can test for the consistency of the bandpowers as well as for the impact of foregrounds at both large and small scales. 
We perform two types of $\ell$-cuts. 
On the low-$\ell$ side, we discard modes with $|\Bell_x| < \ell_{\rm xmin}$ which are mostly affected by the TOD filtering and Galactic dust. 
We apply two $\ell_{\rm xmin}$ cuts, $\ell_{\rm xmin}=50$ and $\ell_{\rm xmin}=200$. 
The largest shift is observed for the $\ell_{\rm xmin}=200$ case where one bandpower is changed by $\approx 1\,\sigma$, although with an uncertainty of $0.6\,\sigma$. 
On the high-$\ell$ side we adjust the maximum multipole value from $\ell_{\rm max}=3000$ to 2500 and 3500. 
This test is sensitive to high-$\ell$ foreground contamination, such as from polarized point sources.
Overall, we find the data are consistent with the expectations from simulations in these $\ell$-cuts tests. \\

\textit{Apodization}: In the baseline analysis we use boundary and point-source masks with a top-hat profile. We test for mask effects by redoing the analysis replacing the baseline mask with one that has been apodized with a cosine profile. Specifically, the cosine taper is set to 10' for the boundary and to 5' for the sources. The induced shift is consistent with expectations based on simulations.

\begin{table}
\caption{Consistency checks\footnote{Results of the consistency checks. For each test we report the $\chi^2$ and PTE of the bandpower-difference as well as the amplitude-difference and the associated PTE.}}
\centering
\begin{tabular}{c | c c | c c }

Test Name & $\chi^2$ & PTE & $\Delta A_{\rm CB} \pm \sigma(\Delta A_{\rm CB})$& PTE\\
\hline
$\ell_{\rm xmin}=50$ & 4.1 & 0.95 & $0.002 \pm 0.033$ & $0.95$\\
$\ell_{\rm xmin}=200$ & 10.1 & 0.45 & $0.001 \pm 0.051$ & $0.99$\\
$\ell_{\rm max}=2500$ & 8.5 & 0.68 & $-0.0005 \pm 0.006$ & $0.94$\\
$\ell_{\rm max}=3500$ & 2.5 & 0.99 & $-0.0003 \pm 0.0013$ & $0.88$\\
Apod. Mask & 9.7 & 0.47 & $-0.020 \pm 0.015$ & $0.23$\\
\hline
\end{tabular}
\label{tab:const_check}
\end{table}
\subsection{Systematic Uncertainties}
In this section we estimate the impact of systematic uncertainties on the measured cosmic rotation power spectrum amplitude.
The sources of systematic uncertainty, as well as their respective impact on $A_{\rm CB}$, are reported in Tab.~\ref{tab:syst_check}.\\

\begin{table}
\caption{Systematic Uncertainties}
\centering
\begin{tabular}{c | c | c}
\hline\hline
Type & $\Delta A_{\rm CB}$ &  $\Delta A_{\rm CB}/\sigma(A_{\rm CB})$ \\
\hline
Beam uncertainty & 0.001 & 0.01 \\
$T/P$ calibration & -0.003 & -0.03 \\
$T \to P$ leakage & -0.002 & -0.02\\
Polarization rotation & -0.0003 & -0.003\\
\hline
\end{tabular}
\label{tab:syst_check}
\end{table}

\textit{Beam uncertainty}: To get a sense of the beam-related systematics we perturb the baseline beam profile using the uncertainties $\Delta F_{\ell}^{\rm beam}$ from \citet{henning18} and convolve the input data maps by $(1 + \Delta F_{\ell}^{\rm beam})$ while leaving the simulations untouched. 
Then, we deconvolve both the data and the simulations with the baseline beam as opposed to $F_{\ell}^{\rm beam}(1 + \Delta F_{\ell}^{\rm beam})$, effectively testing for a systematic 1\,$\sigma$ underestimation of the beam profile over the entire multipole range.
The resulting systematic uncertainty on the CB power spectrum amplitude is $\Delta A_{\rm CB}^{\rm beam}=0.001$, roughly 1\% of the statistical uncertainty on $A_{\rm CB}$. We therefore conclude that the result is robust against beam uncertainty. \\

\textit{Temperature and polarization calibrations}: Errors in the temperature and polarization calibrations will propagate to an uncertainty on the CB power spectrum amplitude; in particular they will affect the reconstructed power spectrum $C_L^{\hat{\alpha}\hat{\alpha}}$ as well as the realization-dependent $N_L^{(0),\rm RD}$ bias.
As discussed in Sec.~\ref{sec:data}, the CMB power measured by \sptpol{} is calibrated to match the \planck observations to better than 1\% accuracy; specifically the $1\,\sigma$ uncertainties on the temperature and polarization calibration factors are $\delta T_{\rm cal}=0.3\%$ and $\delta P_{\rm cal} = 0.6\%$, respectively \citep{henning18}.  
To quantify the impact of these uncertainties we scale the $Q/U$ data maps by $(1+\delta T_{\rm cal})(1 +\delta P_{\rm cal})$ and leave the simulated maps unchanged. 
The difference in the recovered CB amplitudes is $\Delta A_{\rm CB}^{\rm cal} = -0.003$, or $-0.03\,\sigma$, significantly smaller than the statistical uncertainty on $A_{\rm CB}$.  \\

\textit{$T \to P$ leakage}: A mis-estimation of the temperature power leaking into the $Q$ and $U$ maps could also cause a bias in the estimated power spectrum amplitude.
Similarly to the previous systematics, we test for this effect by over-subtracting a $\epsilon^{Q/U}$-scaled copy of the $T$ map by $1\,\sigma$ (in the leakage factors) from the polarization data maps while fixing the rest of the analysis to the baseline case. 
The change induced in $A_{\rm CB}$ is negligibly small at $\Delta A_{\rm CB}^{T\to P} = -0.002$.\\

\textit{Polarization angle rotation}: As already mentioned in Sec.~\ref{sec:data}, there is a 6\% systematic uncertainty in the global orientation of the detectors, which is measured by minimizing the $TB$ and $EB$ correlations. 
The anisotropic CB quadratic estimator is expected to be insensitive to such uncertainty. 
We test for this by rerunning the analysis in the case where we apply an extra 6\% rotation to the data $Q/U$ maps.
We find that $A_{\rm CB}$ shifts by $-0.003\,\sigma$, demonstrating that the bias induced by an offset in the polarization angle rotation is much smaller than statistical uncertainty on the amplitude of the cosmic rotation power spectrum. \\

\subsection{Galactic dust contamination}\label{sec:dust}
At an observing frequency of 150 GHz, the polarized emission from Galactic dust significantly contaminates the $B$-mode signal, especially at large angular scales. 
In this analysis we filter out CMB modes with $|\Bell_x|<100$ before we reconstruct the polarization rotation angle anisotropy, therefore we do not expect significant contamination from Galactic dust, and we checked this in Sec.~\ref{sec:const_checks} by varying the minimum multipole used in the reconstruction process.

To further validate our analysis, and in particular to address the question about the impact of the non-Gaussian dust signature on the recovered cosmic rotation bandpowers, we generate full-sky maps of the polarized dust emission following the scheme outlined in \citet{vansyngel18}. Briefly, this phenomenological model relates the submillimetre polarized thermal dust emission to the structure of the Galactic magnetic field (GMF) and interstellar matter. The GMF is modelled as the sum of a mean uniform field and a Gaussian random turbulent component with a power-law power spectrum, while the structure of interstellar matter is given by the dust total intensity map from \textit{Planck} (we use the GNILC intensity dust map at 353 GHz from \cite{planck18-4}).\footnote{Our non-Gaussian dust simulations include the $E-B$ asymmetry.} The dust realizations match the one-point statistic of the observed polarized fraction over the SPT footprint. The  $Q/U$ dust maps produced at 353 GHz are subsequently scaled to 150 GHz assuming a modified blackbody spectrum for dust with spectral index $\beta_d = 1.53$ and temperature of $T_d=19.6$ K \citep{planck18-11} and then added to our baseline simulations introduced in Sec.~\ref{sec:sims}.

\begin{figure}
	\includegraphics[width=\columnwidth]{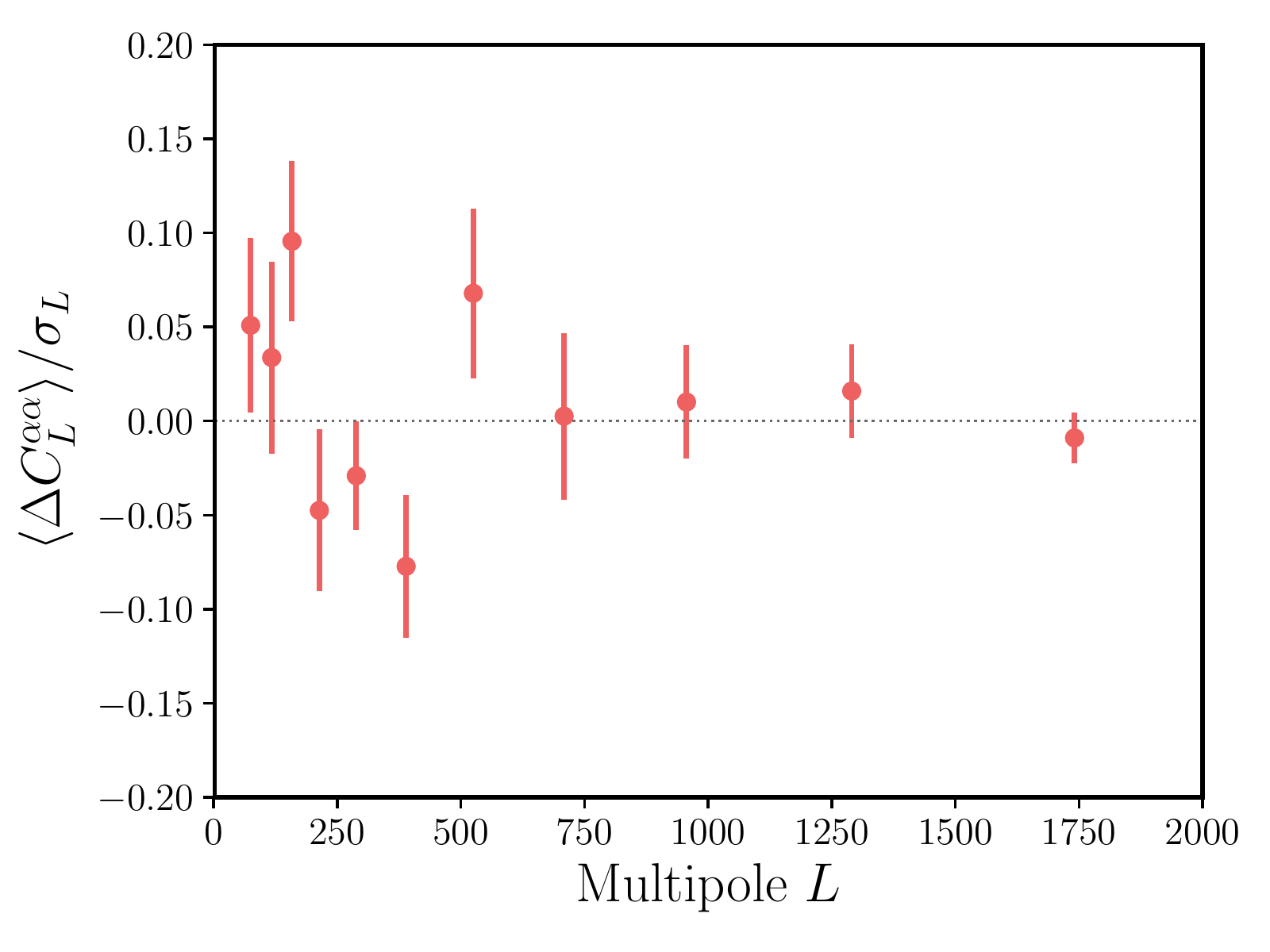}
    \caption{Impact of non-Gaussian polarized Galactic dust. Mean difference cosmic rotation power spectrum between simulations that include the non-Gaussian Galactic dust realizations from \cite{vansyngel18} and the nominal realization with Gaussian foregrounds. The bandpower-difference $\Delta C_L^{\alpha\alpha}$ is normalized by the $1\,\sigma$ statistical uncertainty at each multipole bin. }
    \label{fig:claa_fg_dust}
\end{figure}

In Fig.~\ref{fig:claa_fg_dust} we show the bandpower-difference between simulations that include non-Gaussian dust emission and the baseline ones, averaged over 70 realizations and normalized to the $1\,\sigma$ statistical bandpower uncertainties. As can be seen, the induced shift is at most $0.1\,\sigma$ of the statistical uncertainties at each multipole bin while the PTE under the hypothesis of no difference between the Gaussian and non-Gaussian foregrounds cases is about 15\%. Therefore we conclude that foreground contamination arising from Galactic dust is not significant.

%

%%%%%%%%%%%%%%%%%%%%%%%
%% Results
%%%%%%%%%%%%%%%%%%%%%%%
\section{Results} \label{sec:results}
In this section we present the main results of this analysis: the cosmic rotation power spectrum, the cross-correlation with CMB temperature fluctuations, the scale-invariant CB amplitude, as well as the constraints on two illustrative theoretical models.

\begin{figure*}
\begin{center}
\begin{minipage}{1\linewidth}
\begin{subfigure}{}
\includegraphics[width=0.98\textwidth]{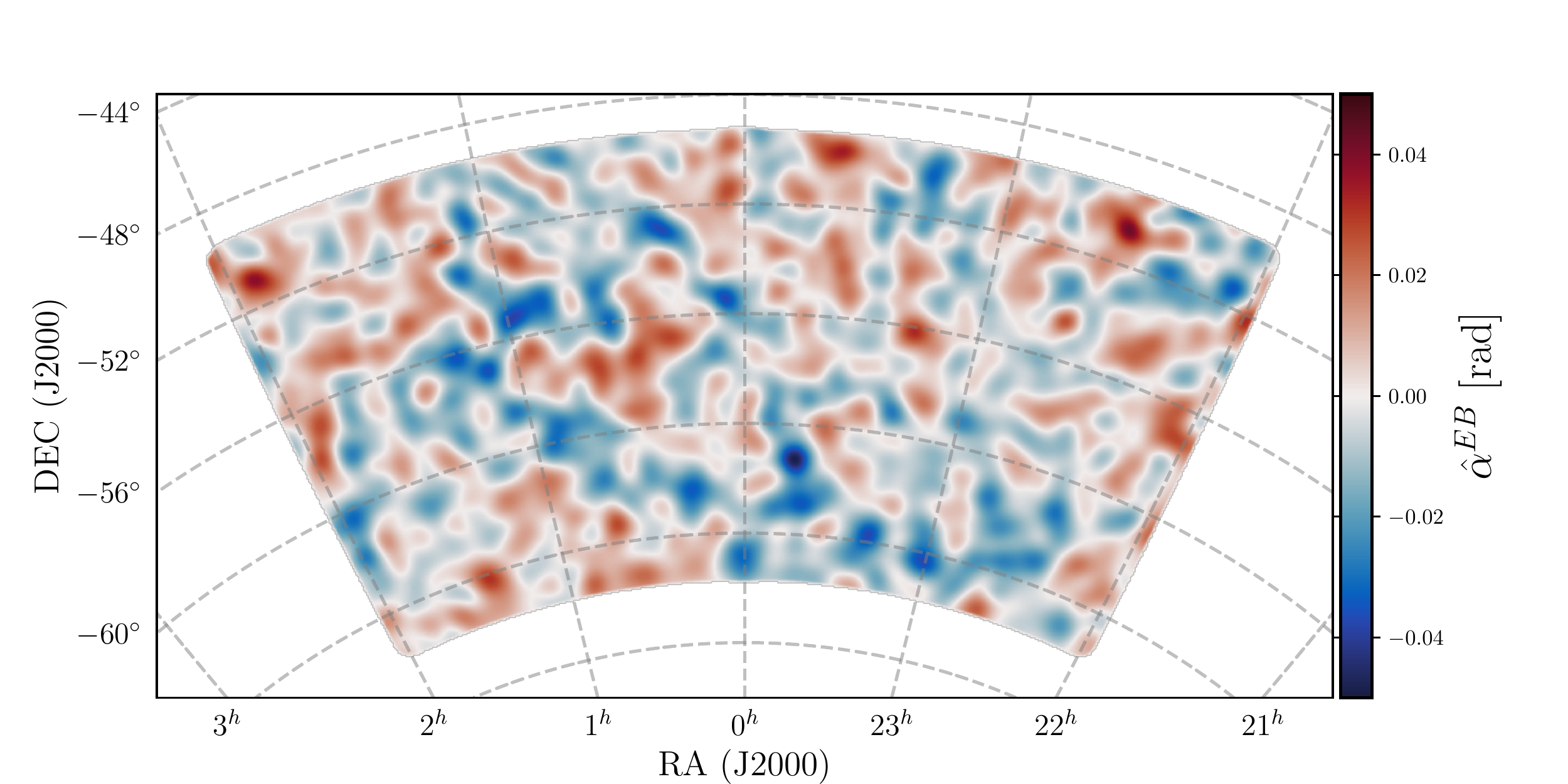}
\end{subfigure}
\end{minipage}
\begin{minipage}{1\linewidth}
\begin{subfigure}{}
\includegraphics[width=0.32\textwidth]{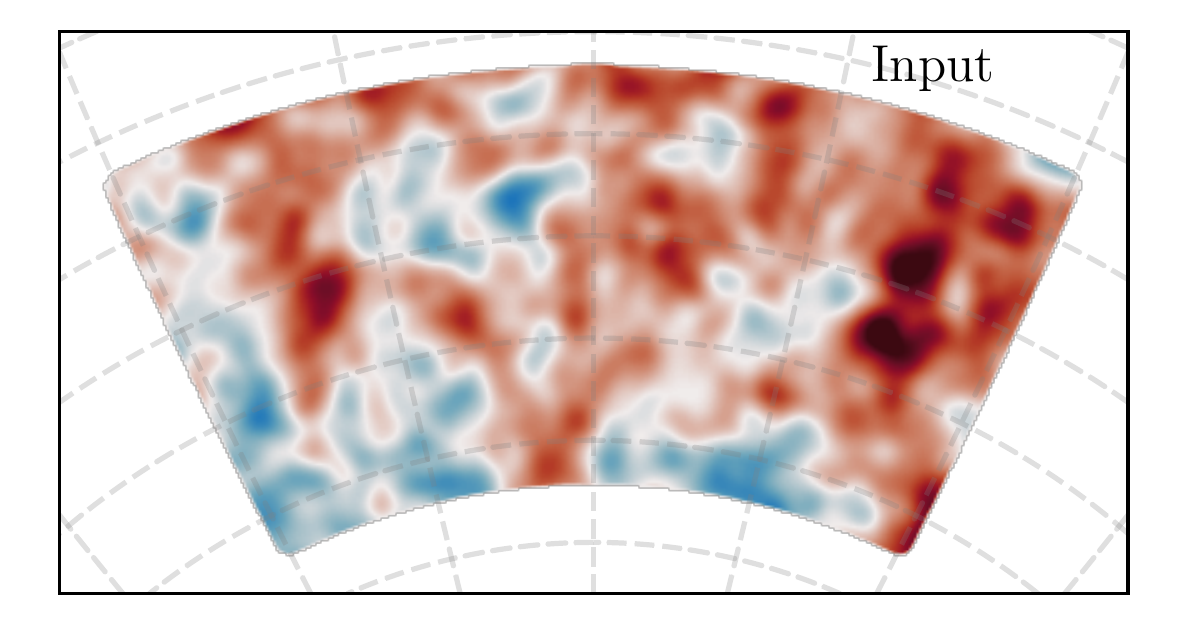}
\includegraphics[width=0.32\textwidth]{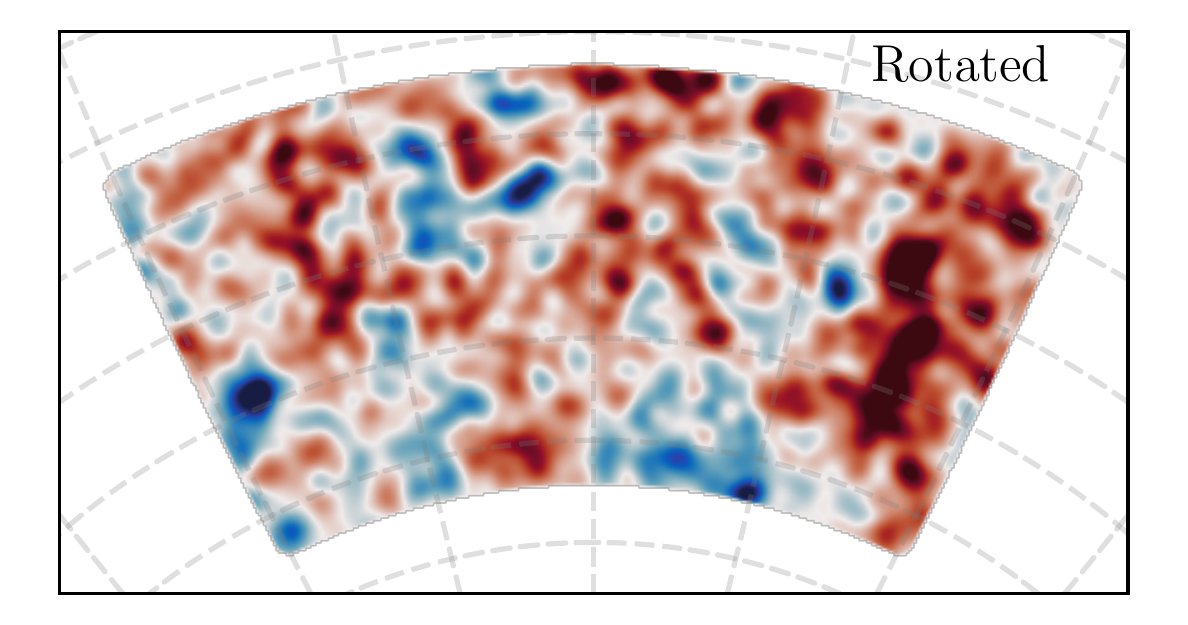} 
\includegraphics[width=0.32\textwidth]{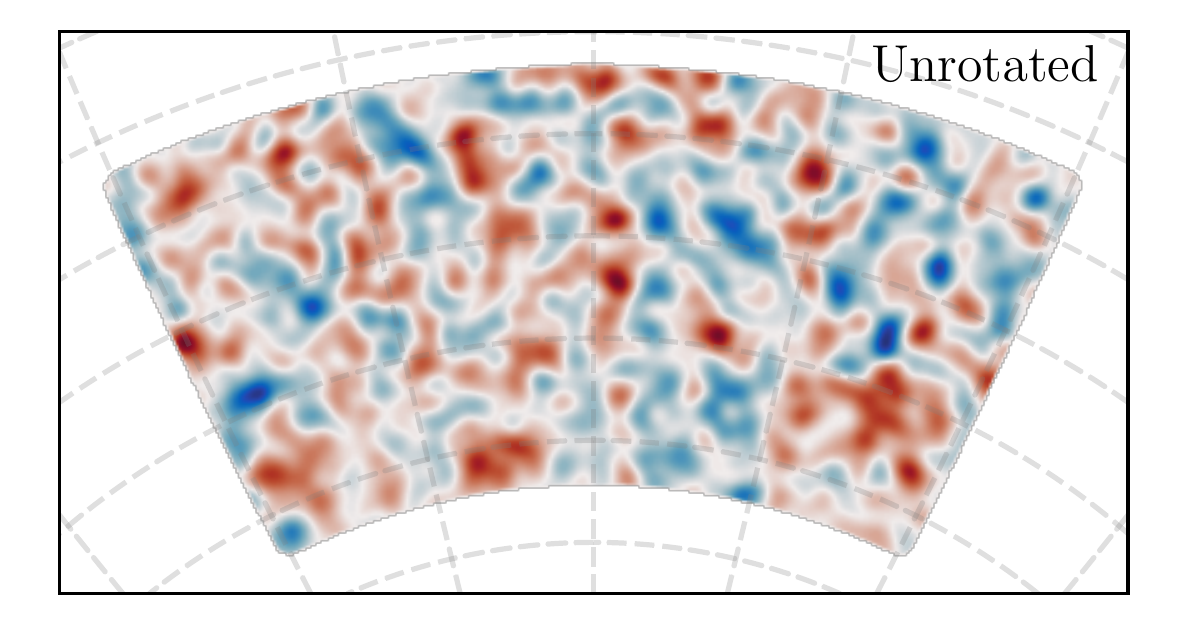} 
\end{subfigure}
\end{minipage}
\caption{{\bf Top}: a map of the reconstructed cosmic birefringence fluctuations $\hat{\alpha}$ from the \sptpol{} 500 \sqdeg{} polarization data using the $EB$ quadratic estimator. The map has been smoothed by a 1 deg FWHM Gaussian beam. {\bf Bottom}: simulated $\alpha$ maps plotted with the same color scale as the top panel and smoothed by a 1 deg FWHM Gaussian beam. The left panel shows the input $\alpha$ map generated from a scale-invariant CB power spectrum with $A_{\rm CB}=1$, the middle panel shows the reconstructed map estimated from the noisy simulation that has been rotated using the input map on the left, and the right panel shows the reconstructed $\alpha$ map obtained from the corresponding unrotated simulation. The pattern of the CB fluctuations reconstructed from the data appears similar to what is seen in the unrotated case, providing a visual indication that the amplitude of the CB signal in the data must be $A_{\rm CB}  \ll 1$.}
\label{fig:alpha_map_data}
\end{center}
\end{figure*}

We start by showing in Fig.~\ref{fig:alpha_map_data} the map of the reconstructed polarization rotation angle fluctuations $\alpha$ over the \sptpol{} 500 \sqdeg{} footprint. 
For visualization purposes the map has been smoothed with a 1 deg FWHM Gaussian kernel.

\subsection{Power spectrum estimation}
\label{sec:results_claa}
\begin{figure}
	\includegraphics[width=\columnwidth]{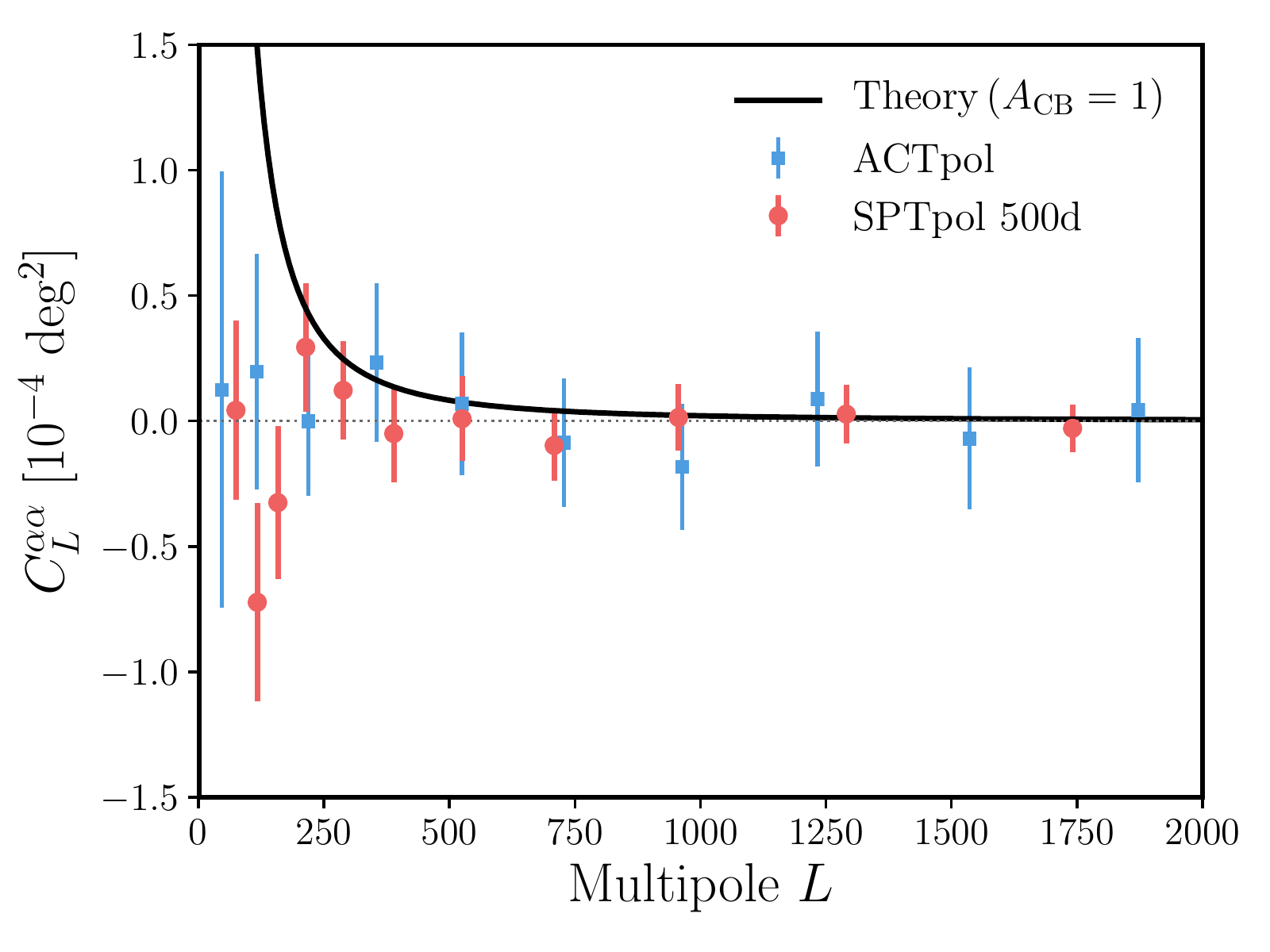}
    \caption{Anisotropic cosmic rotation power spectrum measured from \sptpol{} 500 \sqdeg{} polarization data (red circles) and from the ACTpol experiment \citep[blue squares,][]{namikawa20}. The black solid line shows the fiducial scale-invariant cosmic rotation power spectrum assuming $A_{\rm CB}=1$ (see Eq.~\ref{eq:theory_cl}). The PTE under the no-rotation hypothesis is \pteauto\% and therefore cannot be rejected.}
    \label{fig:claa}
\end{figure}

The cosmic rotation power spectrum measurement from \sptpol{}  is presented in Fig.~\ref{fig:claa}.
We recover the power spectrum in 11  bandpowers in the range $50 \le L \le 2000$.
The bandpower covariance $\mathbb{C}_{L_{b}L_{b'}} $ is estimated using $N_{\rm sim}=300$ simulations of the unrotated skies that have been fully processed through the reconstruction pipeline (see Sec.~\ref{sec:sims}).
The error bars reported are taken from the diagonal of the covariance matrix.
We list in Tab.~\ref{tab:bandpowers} the recovered bandpowers together with their statistical uncertainties.

Our working hypothesis is that the rotation angle map is zero. 
We can calculate the chi-square under this null hypothesis as $\chi^2_{\rm null} = \sum_{bb'} \hat{C}_{L_{b}}^{\alpha\alpha} \mathbb{C}^{-1}_{L_{b}L_{b'}} \hat{C}_{L_{b'}}^{\alpha\alpha} \simeq \chisquareauto$.
The number of simulations with a larger $\chi^2$ than that of the real data translates to a PTE of \pteauto \%, therefore we cannot rule out the no-rotation hypothesis.

\begin{table}
\caption{Cosmic rotation bandpowers from \sptpol{} 500d}
\centering
\begin{tabular}{c c c | c }
\hline\hline
$[\,L_{\rm min}$ & $L_{\rm max}\,]$ & $L_b$ & $ \hat{C}_{L_{b}}^{\alpha\alpha}$ [$\times 10^5$ deg$^2$]\\ [0.5ex]
\hline
$[\,50$&$99\,]$&$75$&$ 0.427\pm3.569$\\                                                                     
$[\,100$&$133\,]$&$117$&$ -7.225\pm3.949$\\                                                                     
$[\,134$&$181\,]$&$158$&$ -3.253\pm3.040$\\                                                                     
$[\,182$&$244\,]$&$213$&$ 2.939\pm2.563$\\                                                                     
$[\,245$&$330\,]$&$288$&$ 1.222\pm1.972$\\                                                                     
$[\,331$&$446\,]$&$389$&$ -0.500\pm1.933$\\                                                                     
$[\,447$&$602\,]$&$525$&$ 0.088\pm1.690$\\                                                                     
$[\,603$&$813\,]$&$708$&$ -0.977\pm1.398$\\                                                                     
$[\,814$&$1097\,]$&$956$&$ 0.140\pm1.328$\\                                                                    
$[\,1098$&$1481\,]$&$1290$&$ 0.274\pm1.174$\\                                                                  
$[\,1482$&$2000\,]$&$1741$&$ -0.293\pm0.948$\\    
\hline
\end{tabular}
\label{tab:bandpowers}
\end{table}

Another way to look at this is by measuring the amplitude of the recovered power spectrum with respect to the fiducial model, as discussed in Sec.~\ref{sec:bin_amp}.
We find an amplitude of the scale-invariant CB power spectrum of $A_{\rm CB} = \ACBamp$, where the statistical uncertainty is derived from the standard deviation of the CB amplitudes from the unrotated simulations.
Finally, note that the results presented in this subsection (as well as in Sec.~\ref{sec:alpha_cross_T}) do not incorporate the marginalization over the estimator's normalization correction $\mathcal{R}_L^{\rm MC}$ but, as mentioned in Sec.~\ref{sec:qe}, this does not bias the power spectrum measurement given the nondetection. 
However, we incorporate the effect of $\mathcal{R}_L^{\rm MC}$ and its uncertainty on the inferred amplitude of the scale-invariant cosmic rotation power spectrum $A_{\rm CB}$ at the likelihood level in Sec.~\ref{sec:constraints}.

\subsection{Cross-correlation with temperature}
\label{sec:alpha_cross_T}
If the CB-inducing field is correlated with primordial density fluctuations, for example in the case of a quintessence field with adiabatic primordial perturbations seeded during inflation, then a cross-correlation signal with CMB temperature fluctuations is also expected \citep[e.g.,][]{caldwell11,capparelli19}.

\begin{figure}
	\includegraphics[width=\columnwidth]{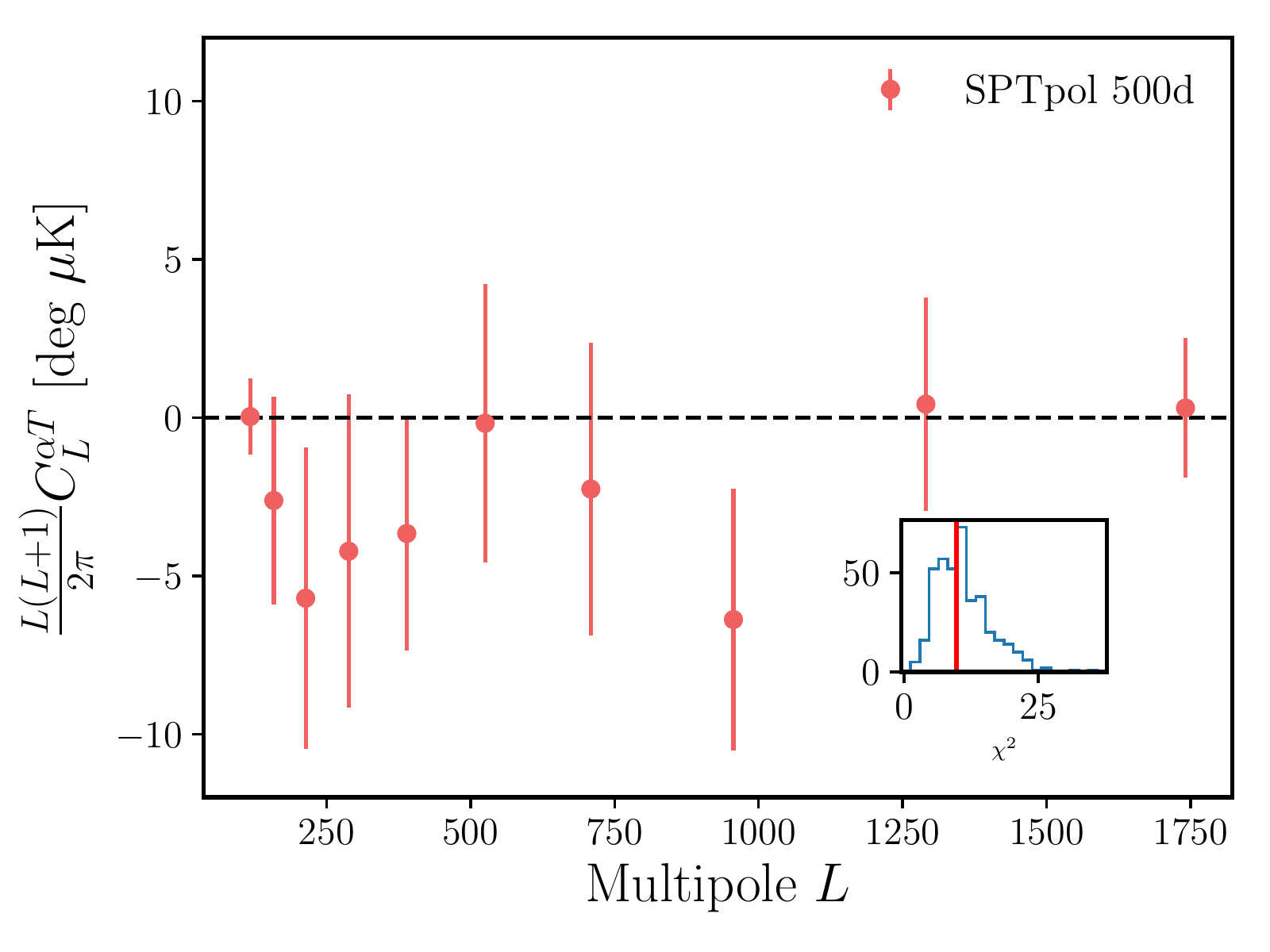}
    \caption{Cross-power spectrum between the \sptpol{} CMB temperature fluctuations and the anisotropic CB angle. The inset panel shows the distribution of $\chi^2_{\rm null}$ from simulations (blue histogram) and the value from data (red vertical line). The cross-power is consistent with the hypothesis of no signal between the maps.}
    \label{fig:claT}
\end{figure}

It is interesting then to cross-correlate the reconstructed rotation angle map $\alpha$ with the CMB temperature fluctuations over the same patch of the sky. 
In Fig.~\ref{fig:claT} we show the cross-spectrum $C_L^{\alpha T}$ reconstructed in 10 bandpowers in the range $100 \le L \le 2000$.
We derive the uncertainties by cross-correlating the simulated temperature and cosmic rotation maps (that have no common cosmological signal) and computing the variance for each bandpower.
Similarly to the auto-spectrum case, we compute the $\chi^2_{\rm null}$ under the no-correlation hypothesis, finding $\chi^2_{\rm null} = \chisquarecross$. 
This corresponds to a PTE of \ptecross\% meaning that, in this case too, we do not reject the null hypothesis.
In addition, the number of simulations with an absolute value of  $\chi_{\rm null} = \sum_{b} C_{L_{b}}^{\alpha T}/\sigma(C_{L_{b}}^{\alpha T})$ larger than that of the data results in a PTE of 16\%.
Despite the reported nondetection, we note that the $C_L^{\alpha T}$ cross-correlation is still informative and can provide tight constraints on the axionlike-photon coupling constant $g_{a\gamma}$ in certain models, even tighter than those provided by cosmic rotation spectrum \citep[e.g,][]{capparelli19}. 
The reason is that while the auto-spectrum  $C_L^{\alpha\alpha}$ depends quadratically on the coupling constant, the cross-spectrum scales as $g_{a\gamma}$ and as such, it is more sensitive to small values of the coupling.

\subsection{Cosmological and fundamental physics implications}
\label{sec:constraints}
The cosmic rotation power spectrum $\hat{C}_L^{\alpha\alpha}$ reconstructed from \sptpol\ data is consistent with the null line.
In order to turn the nondetection into an upper limit on the amplitude of the scale-invariant CB power spectrum $A_{\rm CB}$, we follow the approach of \citet{namikawa20} and construct an approximate likelihood for the recovered CB power spectrum that takes into account small deviations from Gaussianity at the largest scales.
This log-likelihood is based itself on the one proposed by \citet{hamimeche08} and reads

\beq
\label{eq:alpha_like}
-2 \ln{\mathcal{L}_{\alpha}}(A_{\rm CB}) = 
\sum_{bb'} g \left(\hat{A}_{L_b} \right) C^f_{L_b} \mathbb{C}^{-1}_{L_b L_{b'}} C^f_{L_{b'}} g \left(\hat{A}_{L_{b'}} \right),
\eeq
where

\beq
\hat{A}_{L} = \frac{\hat{C}_L^{\alpha\alpha} + N_L^0 + N_L^{\rm lens}}{A_{\rm CB}\left( C_L^{\alpha\alpha} + N_L^1\right) + N_L^0 + N_L^{\rm lens}}
\eeq
is the amplitude of the recovered power spectrum relative to that of simulations including the cosmic birefringence signal $C_L^{\alpha\alpha}$ at a given bin $L_b$, and $g(x) = {\rm sign}(x-1)\sqrt{2(x-\ln{x}-1)}$ for $x \ge 0$.
The fiducial spectrum $C^f_{L}$ and the covariance entering the equation above are measured from the unrotated simulations as discussed in Sec.~\ref{sec:results_claa}.
As mentioned in Sec.~\ref{sec:qe}, we include the effect of a constant multiplicative bias in the response function by rescaling the reconstructed spectrum (as well as the noise biases) according to $\hat{C}_L^{\alpha\alpha} \to \hat{C}_L^{\alpha\alpha}/(\mathcal{R}^{\rm MC})^2$.

We sample the posterior distributions using the \texttt{emcee} package \citep{mackey13} and impose a flat prior on $A_{\rm CB} > 0$, whereas for the normalization factor we adopt the Gaussian prior $P(\mathcal{R}^{\rm MC}) \propto\mathcal{N}(1,0.1^2)$.\footnote{Here $\mathcal{N}(\mu,\sigma^2)$ denotes a  Gaussian distribution with mean $\mu$ and variance $\sigma^2$.}
The resulting 2\,$\sigma$ upper bound on the amplitude of the scale-invariant cosmic rotation power spectrum is $A_{\rm CB} <  \ACBmcmc $, which translates to a limit of $L(L+1)C_L^{\alpha\alpha} / 2\pi < 1.0  \times 10^{-5}$ rad$^2$ (\ACBmcmcsqrad\ \sqdeg{}).\footnote{We note that the 2\,$\sigma$ upper bound on $A_{\rm CB}$ is fairly insensitive to changes in the mean or the variance of the Gaussian prior, such as shifting the mean by $\pm 0.05$ or increasing/decreasing the variance by a factor 2. In particular, if we completely neglect this correction (i.e. we fix $\mathcal{R}^{\rm MC}=1$), we find $A_{\rm CB} < 0.09$.}
This constraint is in line with the $2\,\sigma$ limit  reported by the ACTpol collaboration,  $A_{\rm CB} < 0.1$,  over the multipole range $20 \le L \le 2048$ \citep{namikawa20}. 
As we mentioned in Sec.~\ref{sec:theory}, the largest scales probed by the measurement drive the constraining power, for example if we discard the first bandpower between $50 \le L < 100$ we obtain a 2\,$\sigma$ upper limit of $A_{\rm CB} <  0.15 $. 
Let us finally point out that, as is frequently the case when dealing with upper limits, the specific details of the prior imposed on $A_{\rm CB}$ have a substantial effect on the resulting constraint on the amplitude of the scale-invariant CB power spectrum. 
For instance, adopting the prior $p(A_{\rm CB}) \propto \log{A_{\rm CB}}$ (usually employed when the magnitude of a certain parameter is unknown) results in a 2\,$\sigma$ upper bound of $A_{\rm CB} < 0.026$.
However, the posterior for this prior diverges for small values of $A_{\rm CB}$ and artificially shrinks the inferred upper bounds, as also noted elsewhere in literature \citep[e.g.,][]{polarbear15}.
Therefore, to be more conservative and to facilitate a comparison with previous similar works, we adopt the uniform prior on $A_{\rm CB}$ as our baseline prior.

We can now turn this upper limit into constraints on specific parameters of different physical mechanisms.
Recalling that Eq.~\ref{eq:parity_claa} has been derived under the assumption of an effectively massless pseudoscalar field $a$ at the time of inflation, we can translate the constraint on the scale-invariant cosmic rotation power spectrum to an upper bound on the coupling between axionlike particles and photons,

\beq
g_{a\gamma} \le \frac{\gagammamcmc}{H_I}\quad (95\%\, \text{C.L.}).
 \eeq

This constraint is particularly informative for those models where the axionlike particles have small masses in the $10^{-33} \,{\rm eV} \lesssim m_a \lesssim  10^{-28} \,{\rm eV}$ range.
This mass range can be understood as follows. 
For an axionlike particle with mass $m_a$, the value of $a$ at early times ($H \gg m_a$) is frozen at $a \approx a_0$, while for $H \lesssim m_a$ the field will oscillate around the minimum of its potential, yielding $\Delta a = 0$ (see Eq.~\ref{eq:delta_a}). Therefore, the polarization rotation will be sourced only if the fluctuations of the axionlike field are frozen at recombination and oscillations begin afterwards, i.e. $m_a \lesssim H_{\rm rec} \simeq 10^{-28}$ eV. On the other hand, the mass of the pseudoscalar field has to be large enough for $a$ to be dynamical (i.e. $\dot{a} \ne 0$) between the decoupling and today to produce a polarization rotation. Given that the transition of the field $a$ from static to dynamical occurs when $H \sim m_a$, the lower bound on the mass then becomes $m_a  \gtrsim H_0 \simeq 10^{-33}$ eV.
Considering the current $2\,\sigma$ upper limit on the tensor-to-scalar ratio $r \le 0.07$ \citep{bicep2keck18}, the constraint on the coupling becomes $g_{a\gamma} \le 2.1 r^{-1/2} \times 10^{-16}$ GeV$^{-1}$ $\sim 7.9 \times 10^{-16}$ GeV$^{-1}$ or $6.6 \times 10^{-15}$ GeV$^{-1}$ assuming the forecasted sensitivity $\sigma(r) \simeq 10^{-3}$ from next-generation CMB experiments.

The coupling constant $g_{a\gamma}$ can also be related to the decay constant (or Peccei-Quinn symmetry-breaking scale) $f_a$ through $g_{a\gamma} = (\alpha_{\rm em}/2\pi) \mathcal{C}_{a\gamma}/f_a \sim 10^{-3} /f_a$, where $\alpha_{\rm em}$ is the fine structure constant and $ \mathcal{C}_{a\gamma}$ is a model-dependent dimensionless parameter of $\mathcal{O}(1)$ \citep[e.g.,][]{marsh16}.
Our upper bound on $A_{\rm CB}$ then implies a lower bound on the coupling scale $f_a \gtrsim 4.8 \sqrt{r} \times 10^{12}$ GeV $\sim 1.3 \times 10^{12}$ GeV for $r \sim 0.07$ (or $\sim 1.5 \times 10^{11}$ GeV for $r \sim 10^{-3}$). 
The typical decay constant values predicted in string theory are around the GUT scale, $f_a \sim 10^{16}$ GeV \citep{svrcek06}, and in general below the Planck scale, although values as low as $f_a \sim 10^{10-12}$ GeV are possible \citep{cicoli12}.

Current constraints on the coupling between axionlike particles and photons are based on a wide range of observational and experimental techniques, spanning from astrophysics to terrestrial laboratory experiments. 
For example, the energy loss associated with the production of axions (and other low-mass weakly interacting particles such as neutrinos) inside the interior of globular cluster stars provides a $2\,\sigma$ constraint of $g_{a\gamma} < 6.6 \times 10^{-11}$ GeV$^{-1}$ (or $f_a > 1.5 \times 10^7$ GeV) \citep{ayala14}.
Similarly, helioscopes such as the CERN Axion Solar Telescope (CAST) search for conversions into X-rays of solar axions in a dipole magnet directed towards the Sun and are able to obtain upper bound of $g_{a\gamma} < 6.6 \times 10^{-11}$ GeV$^{-1}$ for $m_a < 0.02$ eV \citep{cast17}.
The absence of $\gamma$-rays from the core-collapse supernova SN1987A, which would originate from the conversion of axionlike particles into photons by the Galactic magnetic field, translates to a constraint of $g_{a\gamma} \lesssim 5.3 \times 10^{-12}$ GeV$^{-1}$ (or $f_a \gtrsim 1.9 \times 10^{8}$ GeV) for $m_a \lesssim 4.4 \times 10^{-10}$ eV \citep{payez15}.
Limits from laboratory searches, such as the Light-Shining-through-Walls or microwave cavity experiments, are currently weaker than astrophysical or cosmological constraints.
For instance, the Optical Search for QED Vacuum Birefringence, Axions, and Photon Regeneration (OSQAR) experiment used a 9\,T transverse magnetic field  and an 18.5 W continuous wave laser emitting at the wavelength of 532 nm to provide a $2\,\sigma$ constraint on $g_{a\gamma} \lesssim 3.5 \times 10^{-8}$ GeV$^{-1}$ (or $f_a \gtrsim 2.9 \times 10^{4}$ GeV) for $m_a \lesssim 0.3$ meV \citep{ballou15}. 

We can also turn the upper limit on $A_{\rm CB}$ into a bound on the strength of a scale-invariant PMF.
Using Eq.~\ref{eq:pmf} and considering an observing frequency of $\nu = 150$ GHz, we find a 95\% upper limit of $B_{1\rm Mpc} < \Bmcmc$ nG.
While current constraints on PMFs from 4-point function measurements like the one presented here are not yet competitive with those from the $B$-mode power spectrum (which are of order $1\,{\text{nG}}$, see, e.g., \citep{zucca17,sutton17}), they will improve dramatically in the near future thanks to the different scalings with $B_{1\rm Mpc}$ \citep{pogosian19}.
In particular, experiments such as CMB-S4 and PICO are projected to obtain bounds on the PMF strength down to $\sim 0.1$ nG, which would rule out the purely primeval origin (without any dynamo mechanism) of the observed $1 - 10 \,\mu$G magnetic fields \citep{grasso01}.
Finally, note that the Faraday rotation caused by a $\sim 0.1$ nG PMF would be similar to that induced by the Galactic magnetic field near the poles \citep{de13}.

%%%%%%%%%%%%%%%%%%%%%%%
%% Conclusions
%%%%%%%%%%%%%%%%%%%%%%%
\section{Conclusions} \label{sec:conclusions}
This paper presents a search for anisotropic cosmic birefringence using CMB polarization data from 500 \sqdeg{} of the sky surveyed with \sptpol{}. 
We apply a quadratic estimator to the observed polarized $E$- and $B$-mode maps and reconstruct a map of the cosmic rotation angle anisotropies.
The amplitude of the recovered power spectrum is consistent with zero.
The 95\% upper limit on the amplitude of the scale-invariant cosmic rotation power spectrum predicted in a wide range of theoretical contexts is $L(L+1)C_L^{\alpha\alpha}/2\pi <  \ACBmcmc \times 10^{-4}$ rad$^2$ (\ACBmcmcsqrad\ \sqdeg{}).
This upper bound is then translated into constraints on the strength of scale-invariant primordial magnetic fields, $B_{1 \rm Mpc} < \Bmcmc \,{\rm nG} $ (95\% C.L.), and on the coupling between axionlike fields and the electromagnetic sector, $g_{a\gamma} \le \gagammamcmc H_I^{-1}$ (95\% C.L.).
We perform a suite of consistency checks and systematic tests to validate the results, finding no evidence for significant contamination.

In addition to the cosmic rotation power auto-spectrum, we have made the first-ever measurement of the cross-correlation between CMB temperature fluctuations and the reconstructed rotation angle map, and find no detectable cosmological signal.

As the instrumental noise level in polarization falls below $\Delta_{P} \approx 5 ~\mu$K-arcmin, the lensed $B$-modes will start dominating the estimator variance, potentially limiting the sensitivity to cosmic birifrigence. 
In principle, delensing techniques \citep[e.g.,][]{manzotti17,polarbear20a} can be applied to the observed $B$-modes to reduce the noise of the estimator to augment the constraining power of the 4-point function estimator \citep{yadav09,pogosian18}. 
More generally, this identical problem arises in CMB lensing, where beyond quadratic estimator techniques have been developed to more optimally extract lensing information from the data, and which could be adapted for cosmic birefringence \citep{carron17,millea19,millea20}. 

Over the next few years the CMB polarization anisotropies will be mapped out over large fractions of the sky with unprecedented sensitivity. 
While the main focus of proposed experiments such as CMB-S4 \citep{cmbs4collab19} and PICO \citep{hanany19} is the detection of primordial tensor perturbations, the data collected will unlock a wide range of ancillary science.
In particular, their promise to improve up to three orders of magnitude the constraints on the amplitude of the scale-invariant cosmic birefringence power spectrum will significantly advance our understanding of primordial magnetism and parity-violating physics \citep{pogosian19}.

%%%%%%%%%%%%%%%%%%%%% Ackn., bib, appendix %%%%%%%%%%%%%%%%%%%%%
\begin{acknowledgements}
The authors would like to acknowledge helpful discussions with Dominic Beck, Giulio Fabbian, Toshiya Namikawa, Giuseppe Puglisi, and Caterina Umilt\`a. 
SPT is supported by the National Science Foundation through grants PLR-1248097 and OPP-1852617.
Partial support is also provided by the NSF Physics Frontier Center grant PHY-1125897 to the Kavli Institute of Cosmological Physics at the University of Chicago, the Kavli Foundation and the Gordon and Betty Moore Foundation grant GBMF 947. This research used resources of the National Energy Research Scientific Computing Center (NERSC), a DOE Office of Science User Facility supported by the Office of Science of the U.S. Department of Energy under Contract No. DE-AC02-05CH11231.  
The Melbourne group acknowledges support from the University of Melbourne and an Australian Research Council's Future Fellowship (FT150100074). 
B.B. is supported by the Fermi Research Alliance LLC under contract no. De-AC02- 07CH11359 with the U.S. Department of Energy.
Work at Argonne National Lab is supported by UChicago Argonne LLC, Operator  of  Argonne  National  Laboratory  (Argonne). Argonne, a U.S. Department of Energy Office of Science Laboratory,  is  operated  under  contract  no.   DE-AC02-06CH11357.  We also acknowledge support from the Argonne  Center  for  Nanoscale  Materials.  
We acknowledge the use of many python packages: \sc{IPython} \citep{ipython}, \sc{matplotlib} \citep{hunter07}, and \sc{scipy} \citep{scipy}.

\end{acknowledgements}
\bibliographystyle{aasjournal}
\bibliography{rot500d.bbl}
%\bibliography{../../../BIBTEX/spt.bib}

\end{document}